\documentclass[letterpaper, 11 pt, onecolumn]{IEEEtran}
\usepackage{epsfig,subfigure,latexsym,amssymb,amsmath}

\newcommand{\eqdef} {\mbox{$\:\stackrel{\triangle}{=}\:$}}
\newtheorem{definition}{Definition}[section]
\newtheorem{remark}{Remark}[section]
\newtheorem{theorem}{Theorem}[section]
\newtheorem{claim}{Claim}[section]

\newtheorem{lemma}{Lemma}[section]
\newtheorem{proposition}[theorem]{Proposition}

\newcommand{\be}{\begin{equation}}
\newcommand{\ee}{\end{equation}}
\newcommand{\ben}{\begin{enumerate}}
\newcommand{\een}{\end{enumerate}}
\newcommand{\bea}{\begin{eqnarray}}
\newcommand{\eea}{\end{eqnarray}}
\newcommand{\bean}{\begin{eqnarray*}}
\newcommand{\eean}{\end{eqnarray*}}

\def\squarebox#1{\hbox to #1{\hfill\vbox to #1{\vfill}}}

\newcommand{\cX}{{\cal X}}
\newcommand{\cY}{{\cal Y}}
\newcommand{\cZ}{{\cal Z}}
\newcommand{\cA}{{\cal A}}

\newcommand{\cB}{{\cal B}}
\newcommand{\cS}{{\cal S}}
\newcommand{\cE}{{\cal E}}

\newcommand{\cI}{{\cal I}}
\newcommand\independent{\protect\mathpalette{\protect\independenT}{\perp}} 
\def\independenT#1#2{\mathrel{\rlap{$#1#2$}\mkern2mu{#1#2}}}

\def\b0{\mathbf{0}}
\def\bX{\mathbf{X}}
\def\bZ{\mathbf{Z}}
\def\bW{\mathbf{W}}

\def\bY{\mathbf{Y}}

\def\tY{\tilde{Y}}

\def\bZ{\mathbf{Z}}

\def\bU{\mathbf{U}}

\def\bW{\mathbf{W}}

\def\bbR{\mathbb{R}}
\def\bbZ{\mathbb{Z}}
\def\bx{\mathbf{x}}
\def\by{\mathbf{y}}

\begin{document}
\baselineskip=1.5\normalbaselineskip

\title{Source and Channel Simulation Using Arbitrary Randomness}
\author{Y\"{u}cel Altu\u{g}~\IEEEmembership{}and Aaron B.~Wagner~\IEEEmembership{Member}
\thanks{The authors are with the School of Electrical and Computer Engineering, Cornell University, 
Ithaca, NY, 14853, USA.
E-mail: {\em ya68@cornell.edu, wagner@ece.cornell.edu}.}}

\maketitle
\begin{abstract}
Necessary and sufficient conditions for approximation of a general channel by a general source are proved. For the special case in which the channel input is deterministic, which corresponds to source simulation, we prove a stronger necessary condition. As the approximation criteria, vanishing variational distance between the original and the approximated quantity is used for both of the problems. Both necessary and sufficient conditions for the two problems are based on some individual properties of the sources and the channel and are relatively easy to evaluate. In particular, unlike prior results for this problem, our results do not require solving an optimization problem to test simulatability. The results are illustrated with several non-ergodic examples. 
\end{abstract}

\section{Introduction}
\label{sec:intro}
The problem of simulating a random process is of fundamental importance in many research disciplines, such as speech processing, applied mathematics, computer science, and the design and performance evaluation of communication systems. Within information theory, one of the motivations is to understand a folklore result which says that `the encoder output of an optimal source code with vanishing error probability is almost uniformly random.' Han \cite{han05} proves that the claim is indeed true when being almost uniformly random is defined as the vanishing normalized divergence distance between the output and the uniform distribution. Gray \cite{gray08} proves the claim for rate-distortion codes and $\bar{d}$-distance, which is defined by Ornstein \cite{ornstein73}. In the literature, there are explicit random number generator constructions based on source codes, of which basic motivation is the aforementioned folklore (cf. \cite{han00, karthik98, gray77} and the references therein). 

The first work on simulation problem dates back at least 40 years, to von Neumann's algorithm to create independent fair coin flips from biased coin flips \cite{vonneumann51} and Elias' \cite{elias72} improvement and fundamental limits. Work within computer science aims to develop `efficient algorithms' (in terms of computational complexity) to simulate random processes (cf. \cite{knuth69,knuth76,devroye86} and references therein). Work within information theory has focused on determining the fundamental limits of the problem in the large-delay limit. In this paper, we follow the information theoretic approach and prove necessary and sufficient conditions for the simulation problem in the following two different setups. These necessary and sufficient conditions are similar to others in information theory in that, while they are not identical, they are similar in essence.  

The first problem we consider is determining whether a general \emph{coin source}\footnote{By general source, we mean a collection of random variables with no consistency requirements among them.} can approximate another general \emph{target source}.\footnote{Here, we adopt the convention used in \cite{han03}, where target (resp. coin) source refers to the to be approximated (resp. to be used for approximation) random source.} By approximation, we mean there should exist a deterministic mapping from realizations of one source to the realizations of the other source, such that the resulting source `well-approximates' the original one in some precise sense. This problem is also known as the \emph{probability distribution approximation problem}.

The second problem is finding necessary and sufficient conditions for determining whether a general \emph{coin source} can approximate a \emph{general channel}, given that another fixed general source is the input to the channel. Similarly, approximation means finding a deterministic mapping from the realizations of the input source and the coin source to the output alphabet of the channel, such that the joint distribution of the input and the simulated output is close to the true distribution. Observe that the channel simulation problem subsumes the source simulation problem, since the latter is a special case of the former with a deterministic input source. 


For the special case of the source simulation problem, for which both coin and target sources are stationary processes with finite states, \cite{elias72} obtains a complete solution for this particular version of the problem. Followup works have proposed more efficient algorithms in terms of computational complexity (cf. \cite{hoeffding70, blum86} and references therein) and$\backslash$or universal algorithms\footnote{By universal algorithm, we mean an algorithm which does not rely on the statistics of the coin source.} (cf. \cite{stout84,perez92} and references therein). 


The case, in which at least one of the sources (or the channel) is non-stationary and non-ergodic, has only received attention more recently. The fundamental work of Han and Verd\'{u} \cite{verdu93}, introduces the  \emph{information spectrum method}, which is the standard tool to handle non-stationary and non-ergodic extensions of the problems in information theory. The basic problem, called \emph{approximation theory of output statistics}, defined in their paper is closely related to the simulation problem, but different from the problems considered in this paper. A somewhat `dual' of this problem, in which the coin source is a general source and the target source is i.i.d. fair coin flips is also solved. For this particular case, Vembu and Verd\'{u} \cite{vembu95} proved necessary and sufficient conditions. Furthermore, Steinberg and Verd\'{u} \cite{steinberg96} considered the problem of simulating a general source using i.i.d. fair coin flips. They proved necessary and sufficient conditions for this problem in the aforementioned work. For the channel simulation problem, Steinberg and Verd\'{u} \cite{steinberg94} proved necessary and sufficient conditions for the special case where the coin source is i.i.d. fair coin flips. All of these works incorporate fundamental notions from the information spectrum method, namely sup-inf entropy rates and sup-inf conditional entropy rates.   

The general case, where both the target random variable and the coin random variable are arbitrary sources, is also investigated in the literature. Results due to Nagaoka \cite{nagaoka96a} (full proofs are also available in \cite{han03}) states a necessary condition and a sufficient condition in terms of sup and inf entropy rates of the target and the coin sources over countable alphabets, however there is a sizable gap between these two conditions, in other words the result is not conclusive. Nagaoka and Miyake \cite{nagaoka96b} state necessary and sufficient conditions without such a gap, for the case of \emph{finite alphabet} sources. However, these conditions are stated in terms of an optimization problem over all joint distributions with the marginals equal to target and coin sources' distribution, hence hard to evaluate.

Our contributions in this paper may be summarized as follows:
\begin{itemize}
\item We state new necessary and sufficient conditions, which are essentially the same, for a coin source to be an approximating source of a given target source, where both of the sources are general sources over countable alphabets. Our necessary condition is strictly stronger than its state-of-the-art counterpart stated in \cite{nagaoka96b}.

\item We state the first, to the best of our knowledge, necessary and sufficient conditions, which are essentially the same, for a coin source to be an approximating source of a channel, given a fixed input source to the channel, where both coin and input sources and the channel are general ones over countable alphabets.  

\item Unlike the existing conditions in the literature, our conditions do not include a $1/n$ scaling factor.
\end{itemize}

Note that both necessary and sufficient conditions for both of the problems are in terms of the intrinsic properties of the sources and the channel, in which the whole spectrum is exploited, as opposed to the traditional quantities like entropy and conditional entropy rates, which are the limiting points of the entropy and conditional entropy spectrum, respectively. Hence, this kind of approach may lead to solutions to some of the open problems of the non-stationary and non-ergodic information theory, such as \cite{karthik00}.

The paper is organized as follows. In Section~\ref{sec:notation-definitions} we state our notation used throughout the paper, give basic definitions and state our results. We also include examples to illustrate our results. Section~\ref{sec:source-approximation} is devoted to the proof of sufficient and necessary conditions for source simulation problem and the demonstration of the fact that our necessary condition is strictly stronger than its state-of-the-art counterpart, while Section~\ref{sec:channel-approximation} consists of the proof of sufficient and necessary conditions for channel simulation problem. The paper ends with conclusions, stated in Section~\ref{sec:conclusion}.

\section{Notation, Definitions and Statement of the Results, Examples}
\label{sec:notation-definitions}

\subsection{Notation}
\label{ssec:notation}
Boldface letters denote vectors; regular letters with subscripts denote individual elements of vectors. Furthermore, capital letters represent random variables and lowercase letters denote individual realizations of the corresponding random variable. Throughout the paper, all logarithms are base-$e$, unless otherwise specified. For $p \in [0,1]$, $H(p)$ denotes the binary entropy function. $\overline{\bbR} \eqdef \bbR \cup \{ -\infty, +\infty \}$ denotes extended real numbers. $(\bbR, \cB, \mu)$ denotes a measure space, with $\cB$ denoting the Borel-sigma algebra on real numbers and $\mu$ denoting the Lebesgue measure. For an arbitrary sequence of real-valued random variables $\{ Z_n\}_{n=1}^\infty$, $p-\liminf_{n \rightarrow \infty} Z_n \eqdef \{ \alpha \, : \, \lim_{n \rightarrow \infty} \Pr\{ Z_n > \alpha\} = 0 \}$ denotes the ``limit infimum in probability'', (cf. Definition 1.3.1. of \cite{han03}). $\Omega \sim U[0,1]$ is a shorthand notation for ``$\Omega$ is a uniform random variable over $[0,1]$''. Given a random variable $X$ with p.m.f. $P_X$, $\mbox{E}_{P_X}[\cdot]$ denotes expectation with respect to $P_X$.

\subsection{Definitions and Statement of the Results}
\label{ssec:definitions}

\begin{definition}
Given two random variables $X,Y \in \cX$, such that $\cX$ is countable set, with pmfs $P_X$ and $P_Y$, respectively; \emph{the variational distance between $P_X$ and $P_Y$}, denoted by $d(P_X,P_Y) $, is defined:
\begin{equation}
d(P_X,P_Y) \eqdef \sum_{x \in \cX} |P_X(x) - P_Y(x)|.
\label{eq:def-var-dist}
\end{equation}
Note that we will also use $d(X,Y)$ to denote the variational distance throughout the rest of the paper, interchangeably with $d(P_X,P_Y)$ to denote the quantity in \eqref{eq:def-var-dist}.
\label{def:var-dist}
\end{definition}

\begin{definition}
Let $\bX = \left\{ X_n \right\}_{n=1}^{\infty}$ and $\bY = \left\{ Y_n \right\}_{n=1}^{\infty}$ be two general sources, where for all $n \in \bbZ^+$, $X_n$ and $Y_n$ are random variables taking values in $\cX_n$ and $\cY_n $, respectively, such that $\cX_n$ and $\cY_n$ are countable sets. We say that \emph{$\bX$ is an approximating source for $\bY$}, if there exists a sequence of deterministic mappings\footnote{All of the mappings, which are mentioned throughout the rest of the paper, are deterministic ones and for the sake of convenience, we drop the quantifier `deterministic' from now on.} $\{ \phi_n : \cX_n \rightarrow \cY_n\}_{n=1}^{\infty}$, such that $\lim_{n \rightarrow \infty} d\left( Y_n, \phi_n\left(X_n\right)\right) = 0$.
\label{def:appr-source}
\end{definition}

\begin{definition}
Let $\bX = \left\{ X_n \right\}_{n=1}^{\infty}$ and $\bZ = \left\{ Z_n \right\}_{n=1}^{\infty}$ be arbitrary general sources, where for all $n \in \bbZ^+$, $X_n$ and $Z_n$ are random variables taking values in $\cX_n$ and $\cZ_n $, respectively, and $\cX_n$ and $\cZ_n$ are countable sets with a given coupling\footnote{For any pair of random variable such a coupling exists, i.e. product distribution of marginals gives a joint distribution of $X_n$ and $Y_n$. In fact, as far as practical application goes, this case is the most interesting case.} $P_{Z_n|X_n}$ between them. Let, $\bW_{Y|X} = \left\{ W_{Y_n|X_n}(Y_n|X_n) \right\}_{n=1}^{\infty}$ be a general channel, where for all $n \in \bbZ^+$, $W_{Y_n|X_n}$ denotes a conditional pmf over $\cY_n \times \cX_n$, where $\cY_n$ is countable set.  We say that \emph{$\bZ$ is an approximating source for $\bW_{Y|X}$, given $\bX$}, if there exists a sequence of mappings $\{ \varphi_n : \cX_n \times \cZ_n \rightarrow \cY_n\}_{n=1}^{\infty}$, such that $\lim_{n \rightarrow \infty} d\left(X_nY_n, X_n \varphi_n(X_n,Z_n) \right) = 0$.
\label{def:appr-source-channel}
\end{definition}

We state our necessary and sufficient conditions in terms of the following quantity.
\begin{definition}
Given a random variable $Z_n$ taking values in $\cZ_n$, where $\cZ_n$ is countable set, let $\cS_n(Z)$ denote an ordered list of $z \in \cZ_n$ sequences with non-zero probability, from highest probable to lowest probable, i.e. 
\begin{equation}
\cS_n(Z) \eqdef  \{ z_i \}_{i=1}^{\infty}, \mbox{ s.t. } P_{Z_n}(z_1) \geq \ldots \geq P_{Z_n}(z_{i}) \geq \ldots. 
\label{eq:defn-sz}
\end{equation}
Next, using $\cS_n(Z)$, define the following partition of $[0,1]$
\begin{equation}
\Delta \eqdef \{ 0 = \delta_0 < \delta_1 < \ldots \leq 1\},
\label{eq:defn-c-partition}
\end{equation}
such that $\forall i \in \bbZ^+, \, \delta_{i} - \delta_{i-1} = P_{Z_n}(z_i)$. For any $\delta \in [0,1)$, 
\begin{equation}
c_n^z(\delta) \eqdef \log \frac{1}{P_{Z_n}(z_{k})},
\label{eq:defn-c}
\end{equation}
such that $\delta \in [\delta_{k-1}, \delta_k)$ for some $k \in \bbZ^+$ by using \eqref{eq:defn-c-partition}. Note $c_n^z(\delta)$ is a well-defined quantity.
\label{def:c-n}
\end{definition}

\begin{figure}[!htb]
\centering \centerline{\epsfxsize = 5.0in
\epsfbox{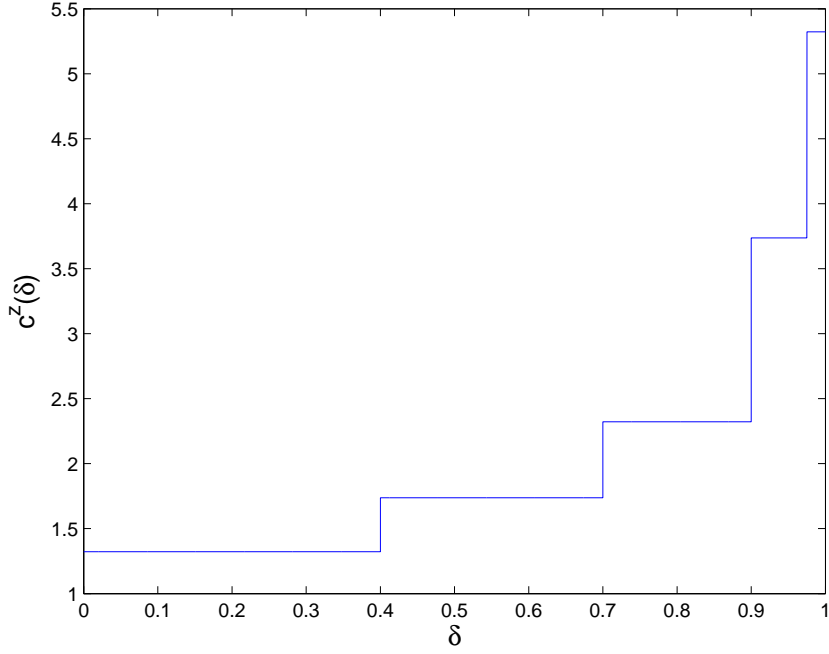}} 
\caption{Graphical representation of $c^z(\cdot)$ for $Z \in \{ z_1, z_2, z_3, z_4, z_5\}$ with the distribution $P_Z(z_1) = 0.025$, $P_Z(z_2) = 0.075$, $P_Z(z_3) = 0.2$, $P_Z(z_4) = 0.3$ and $P_Z(z_5) = 0.4$.} 
\label{fig:fig1}
\end{figure}

Observe that, $c_n^z(\delta) \geq 0$, for all $n$ and for all $\delta \in [0,1)$. Further, $c_n^z(\delta)$ is non-decreasing and right continuous in $\delta$.

\begin{remark}
Throughout the rest of the paper, when we refer to a quantity including $\mu$ (which stands for Lebesgue measure), it should be explicitly understood that we are using $(\bbR, \cB, \mu)$ as our measure space.
\end{remark}

Next, we state our main results:
\begin{enumerate}
\item{\textbf{Source Approximation:}} Consider $\bX = \{ X_n\}_{n=1}^\infty$ and $\bY = \{ Y_n\}_{n=1}^\infty$, where for all $n \in \bbZ^+$, $X_n$ is a random variable taking values in $\cX_n$, such that $\cX_n$ is countable set  (resp. $Y_n$ is a random variable taking values in $\cY_n$, such that $\cY_n$ is countable set).\\
\underline{Sufficient Condition}: If 
\begin{equation}
\forall \gamma \in \bbR,  \, \lim_{n \rightarrow \infty} \mu \left( \delta \in [0,1) \, : \, c_n^x(\delta) - c_n^y(\delta) < \gamma \right) = 0,
\label{eq:suffi-cond-intro}
\end{equation}
then $\bX$ is an approximating source for $\bY$. Note that, \eqref{eq:suffi-cond-intro} can also be written as
\begin{equation}
\mu - \liminf_{n \rightarrow \infty} \{ c_n^x(\delta) - c_n^y(\delta) \} = \infty,
\end{equation}
where $\mu - \liminf$ is the analogous of $p - \liminf$ quantitiy using Lebesgue mesaure instead of the probability measure.
\\
\underline{Necessary Condition}: If $\bX$ is an approximating source for $\bY$, then
\[ 
\inf_{0 < \epsilon < 1} \liminf_{n \rightarrow \infty} \inf_{0 \leq \delta < 1 - \epsilon} \{ c_n^x(\delta + \epsilon) - c_n^y(\delta) \} \geq 0.
\]
\item{\textbf{Channel Approximation:}} Consider $\bX = \{ X_n\}_{n=1}^\infty$ and $\bZ = \{ Z_n\}_{n=1}^\infty$, where for all $n$, $X_n$ is a random variable taking values in $\cX_n$, such that $\cX_n$ is countable set  (resp. $Z_n$ is a random variable taking values in $\cZ_n$, such that $\cZ_n$ is countable set) with a coupling $P_{Z_n|X_n}$. Further, $\bW_{Y|X} = \left\{ W_{Y_n|X_n}(Y_n|X_n) \right\}_{n=1}^{\infty}$, where for all $n \in \bbZ^+$, $W_{Y_n|X_n}$ denotes a conditional pmf over $\cY_n \times \cX_n$, with $\cY_n$ being countable set. For all $n \in \bbZ^+$, for any $x \in \cX_n$ and $\delta \in [0,1)$, $c_n^{z|x}(\delta, x)$ (resp. $c_n^w(\delta, x)$) denotes the quantity defined in Definition~\ref{def:c-n} for $P_{Z_n|X_n}(\cdot | x)$ (resp. $W_{Y_n|X_n}(\cdot | x)$).\\
\underline{Sufficient Condition}: If
\[ 
\forall \gamma \in \bbR, \, \lim_{n \rightarrow \infty} \mbox{E}_{P_{X_n}}\left[\mu( \delta \in [0,1) \, : \, c_n^{z|x}(\delta, X_n) - c_n^w(\delta, X_n) < \gamma) \right] = 0,
\]
then $\bZ$ is an approximating source for $\bW$, given $\bX$. \\
\underline{Necessary Condition}: If $\bZ$ an approximating source for $\bW$, given $\bX$, then
\[
\forall \epsilon \in (0,1), \, \forall \gamma \in \bbR^+, \, \lim_{n \rightarrow \infty} \mbox{E}_{P_{X_n}}\left[\mu( \delta \in [0,1-\epsilon) \, : \, c_n^{z|x}(\delta + \epsilon, X_n) - c_n^w(\delta, X_n) <  - \gamma) \right] = 0.
\]
\end{enumerate}
\begin{remark}
For the sake of comparison, now we state the necessary and sufficient conditions stated in \cite{nagaoka96b}. Let $\cX_n$ (resp. $\cY_n$) be a set with $|\cX_n| < \infty$ (resp. $|\cY_n| < \infty$) for all $n \in \bbZ^+$ and $X_n$ (resp. $Y_n$) be a random variable defined over $\cX_n$ (resp. $\cY_n$) with distribution $P_{X_n}$ (resp. $P_{Y_n}$). Define $\bX = \{ X_n\}_{n=1}^\infty$ and $\bY = \{ Y_n\}_{n=1}^\infty$ and $\mathcal{P}(X_n, Y_n) \eqdef \left\{ \{ P_{X_n, Y_n}(X_n, Y_n) \}_{n=1}^{\infty} \, : \, \mbox{ The marginal distributions are } P_{X_n} \mbox{ and } P_{Y_n}, \, \forall n \in \bbZ^+ \right\}$.
\begin{itemize}
\item If 
\begin{equation}
\sup_{\{ P_{X_n, Y_n}(X_n, Y_n) \}_{n=1}^{\infty} \in \mathcal{P}(X_n,Y_n)} p-\liminf_{n \rightarrow \infty}\left\{ \frac{1}{n} \log \frac{1}{P_{X_n}(X_n)} - \frac{1}{n}\log \frac{1}{P_{Y_n}(Y_n)}\right\} >0,
\label{eq:japanese-sufficient}
\end{equation}
then $\bX$ is an approximating source for $\bY$. 
\item If $\bX$ is an approximating source for $\bY$, then 
\begin{equation}
\sup_{\{ P_{X_n, Y_n}(X_n, Y_n) \}_{n=1}^{\infty} \in \mathcal{P}(X_n,Y_n)} p-\liminf_{n \rightarrow \infty}\left\{ \frac{1}{n} \log \frac{1}{P_{X_n}(X_n)} - \frac{1}{n} \log\frac{1}{P_{Y_n}(Y_n)}\right\} \geq 0.
\label{eq:japanese-necessary}
\end{equation}
\end{itemize}

We demonstrate in Section~\ref{ssec:source-approximation-comparison} (cf. Remark~\ref{rem:remark-necc-cond-comp2}) that our necessary condition is strictly stronger than the one given in (8). Moreover, our conditions do not involve an optimization problem, and hence are easier to evaluate, as the following examples show.
\end{remark}


\subsection{Examples}
\label{ssec:examples}
In this section we provide the following examples to illustrate the necessary and sufficient conditions for source and channel simulation.

\textbf{Example 1:} Let $\{ \bX_1^n\}_{n \geq 1}$, $\{ \bX_2^n \}_{n \geq 1}$, $\{ \bY_1^n \}_{n \geq 1}$ and $\{ \bY_2^n \}_{n \geq 1}$ be stationary, memoryless Bernoulli sources with parameters $p_1$, $p_2$, $q_1$ and $q_2$, respectively. Observe that we have \cite{cover91}
\begin{eqnarray}
& & \forall i \in \{ 1, 2 \}, \, \frac{1}{n}\log \frac{1}{P_{X_i}(\bX^n)} \rightarrow H(p_i ) \, \mbox{(a.s.) as } n \rightarrow \infty, \label{eq:example1-1}\\
& & \forall i \in \{ 1, 2 \}, \, \frac{1}{n}\log \frac{1}{P_{Y_i}(\bY^n)} \rightarrow H(q_i ) \, \mbox{(a.s.) as } n \rightarrow \infty. \label{eq:example1-2}
\end{eqnarray}
Next, we define coin and target sources as mixtures of these in the following way:
\begin{align}
\forall n \in \bbZ^+, \, \bX^n & \eqdef Q_1 \bX_1^n + (1 - Q_1) \bX^n_2, \label{eq:example1-3}\\
\forall n \in \bbZ^+, \, \bY^n & \eqdef Q_2 \bY_1^n + (1 - Q_2) \bY^n_2, \label{eq:example1-4}
\end{align}
where $Q_1$ (resp. $Q_2$) denotes an independent Bernoulli random variable with parameter $\alpha \in [0,1/2]$ (resp. $\beta \in [0,1/2]$). Let $\bX = \left\{ \bX^n \right\}_{n \geq 1}$  (resp. $\bY = \left\{ \bY^n \right\}_{n \geq 1}$) denotes the coin (resp. target) source. Suppose $ \forall n \in \bbZ^+, \, Q_1 \independent \bX_1^n \independent \bX^n_2$ and $Q_2 \independent \bY_1^n \independent \bY^n_2$. Further, suppose that we have
\begin{equation}
0 < q_1 < p_1 < q_2 < p_2 < 1/2.
\label{eq:example1-6}
\end{equation}
Using \eqref{eq:example1-1}, \eqref{eq:example1-2}, \eqref{eq:example1-3} and \eqref{eq:example1-4} we have
\begin{equation}
\lim_{n \rightarrow \infty} \left\{ \frac{c_n^x(\delta)}{n} - \frac{c_n^y(\delta)}{n} \right\} = H(p_1) \mathbf{1}_{\{0 < \delta < \alpha\}} + H(p_2) \mathbf{1}_{\{\alpha < \delta < 1\}} - H(q_1) \mathbf{1}_{\{0 < \delta < \beta\}} - H(q_2) \mathbf{1}_{\{\beta < \delta < 1\}},
\label{eq:example1-5}
\end{equation} 
for any $\delta \in [0,1)$, such that $\delta \neq \alpha$ and $\delta \neq \beta$. 

Next, we consider the following cases for $\alpha$ and $\beta$.
\begin{enumerate}
\item\underline{$\alpha \leq \beta$}: Using \eqref{eq:example1-6} and \eqref{eq:example1-5} we have
\begin{equation}
\forall \delta \in [0,1), \mbox{ s.t. } \delta \neq \alpha, \, \delta \neq \beta \, \lim_{n \rightarrow \infty} \left\{ \frac{c_n^x(\delta)}{n} - \frac{c_n^y(\delta)}{n} \right\} \geq \min\{ H(p_2) - H(q_2), H(p_1) - H(q_1)\} > 0.
\label{eq:example1-7}
\end{equation}
\eqref{eq:example1-7} immediately implies that
\begin{equation*}
\exists \gamma \in \bbR^+, \mbox{ s.t. } \lim_{n \rightarrow \infty} \mu\left( \delta \in [0,1) \, : \, \frac{c_n^x(\delta)}{n} - \frac{c_n^y(\delta)}{n} < \gamma \right) = 0,
\end{equation*}
which implies that
\begin{equation}
\forall \gamma \in \bbR^+, \, \lim_{n \rightarrow \infty} \mu(\delta \in [0,1) \, : \, c_n^x(\delta) - c_n^y(\delta) < \gamma) =0.
\label{eq:example1-8}
\end{equation}
Recalling the sufficient condition for source approximation, \eqref{eq:example1-8} implies that $\bX$ is an approximating source for $\bY$. 
\item\underline{$\alpha > \beta$}: Define $ \bbR^+ \ni \epsilon \eqdef  \frac{\alpha-\beta}{2}$ and consider any $\delta \in \left( \beta ,\frac{\alpha}{2} + \frac{\beta}{2} \right)$. Recalling \eqref{eq:example1-6} and \eqref{eq:example1-5}, this implies

\begin{equation*}
\lim_{n \rightarrow \infty} \left\{ \frac{c_n^x(\delta + \epsilon)}{n} - \frac{c_n^y(\delta)}{n} \right\} = H(p_1) - H(q_2) < 0,
\end{equation*}
which immediately implies that
\begin{equation}
\inf_{\epsilon \in (0,1)} \liminf_{n \rightarrow \infty} \inf_{\delta \in [0, 1- \epsilon)} \left\{ \frac{c_n^x(\delta + \epsilon)}{n} - \frac{c_n^y(\delta)}{n} \right\} < 0.
\label{eq:example1-9}
\end{equation}
Using \eqref{eq:example1-9}, we conclude that
\begin{equation}
\inf_{\epsilon \in (0,1)} \liminf_{n \rightarrow \infty} \inf_{\delta \in [0, 1- \epsilon)} \left\{ c_n^x(\delta + \epsilon) - c_n^y(\delta) \right\} < 0.
\label{eq:example1-10}
\end{equation}
Recalling the necessary condition for source approximation, \eqref{eq:example1-10} implies that $\bX$ is \emph{not} an approximating source for $\bY$. 
\end{enumerate}

\textbf{Example 2:} Let $\{ \bX_1^n\}_{n \geq 1}$, $\{ \bX_2^n \}_{n \geq 1}$, $\{ \bZ_1^n \}_{n \geq 1}$ and $\{ \bZ_2^n \}_{n \geq 1}$ be stationary, memoryless Bernoulli sources with parameters $r_1$, $r_2$, $p_1$ and $p_2$, respectively. Define
\begin{align}
\forall n \in \bbZ^+, \, \bZ^n & \eqdef Q_1 \bZ_1^n + (1 - Q_1) \bZ^n_2, \label{eq:example2-1} \\
\forall n \in \bbZ^+, \, \bX_n & \eqdef Q_2 \bX_1^n + (1 - Q_2) \bX^n_2, \label{eq:example2-2}
\end{align}
where $Q_1$ (resp. $Q_2$) is Bernoulli random variable with parameter $\alpha \in [0,1/2]$ (resp. $\theta \in [0,1/2]$). Let $\bX = \left\{ \bX^n \right\}_{n = 1}^\infty$  (resp. $\bZ = \left\{ \bZ^n \right\}_{n = 1}^\infty$) denotes the input (resp. coin) source. Suppose $\forall n \in \bbZ^+, \, Q_1, Q_2, \bX_1^n, \bX^n_2, \bZ_1^n, \bZ_2^n$ are independent. Further, we define our channel in the following way
\begin{equation}
W_{Y_n|X_n} = \left\{ \begin{array}{cl} W^1_{Y^n|X^n}, & \mbox{ with probability }\beta,   \\
W^2_{Y^n|X^n}, & \mbox{ with probability } 1 - \beta,\\
\end{array} \right.
\label{eq:example2-3}
\end{equation}
where for $i \in \{ 1,2\}$, $W^i_{Y^n|X^n}(\by^n|\bx^n) = \prod_{j=1}^nW^i_{Y|X}(y_j|x_j)$, and $W^i_{Y|X}$ is a BSC with crossover probability $q_i \in (0, 1/2)$ and $\bW_{Y|X} = \{ W_{Y^n|X^n}\}_{n=1}^\infty$.
Equivalently, we can define $\bY^n$, output of the channel due to the input $\bX^n$, in the following way. First, let $\{ \bU_1^n \}_{n \geq 1}$ and $\{ \bU_2^n \}_{n \geq 1}$ denote stationary, memoryless Bernoulli sources with parameters $q_1$ and $q_2$, respectively. Then, for any $n \in \bbZ^+$, define $\bY_1^n \eqdef \bX^n \oplus \bU_1^n$ and $\bY_2^n \eqdef \bX^n \oplus \bU_2^n$. Hence, we have
\begin{equation}
\bY^n \eqdef Q_3 \bY^n_1 + (1 - Q_3) \bY^n_2,
\label{eq:example2-4}
\end{equation}
where $Q_3$ is a Bernoulli random variable with parameter $\beta$. Suppose $\forall n \in \bbZ^+, \, Q_1 \independent Q_2 \independent Q_3 \independent \bX_1^n \independent \bX^n_2 \independent \bZ_1^n \independent \bZ_2^n \independent \bU_1^n \independent \bU_2^n$. Further, suppose that we have
\begin{equation}
0 < q_1 < p_1 < q_2 < p_2 < 1/2.
\label{eq:example2-8}
\end{equation}
Observe that for any $\bx^n \in \{ 0,1 \}^n$, $\frac{1}{n}\log\frac{1}{W^i_{Y^n|X^n}(\bY^n|\bx^n)} = \frac{1}{n}\log\frac{1}{P_{U_i^n}(\bU_i^n)}$, $\forall i \in \{1,2\}$. Using this along with \eqref{eq:example2-4}, and recalling \eqref{eq:defn-c} (observe the independence of $\bZ$ and $\bX$), we have
\begin{equation}
\lim_{n \rightarrow \infty} \left\{ \frac{c_n^{z|x}(\delta, \bx^n)}{n} - \frac{c_n^w(\delta, \bx^n)}{n} \right\} =  H(p_1) \mathbf{1}_{\{ 0 < \delta < \alpha \}} + H(p_2) \mathbf{1}_{\{ \alpha < \delta < 1 \}} - H(q_1) \mathbf{1}_{\{ 0 < \delta < \beta \}} - H(q_2) \mathbf{1}_{\{ \beta < \delta < 1 \}},
\label{eq:example2-7}
\end{equation}
for all $\{ \bx^n\}_{n=1}^\infty$ and for all $\delta \in [0,1)$ except $\delta = \alpha$ and $\delta = \beta$.

Next, we consider the following cases for $\alpha$ and $\beta$.
\begin{enumerate}
\item \underline{$\alpha \leq \beta$}: First of all, observe that using \eqref{eq:example2-8}, we have
\begin{equation}
\exists M \in \bbR^+, \mbox{ s.t. } M < \min\{ H(p_2)-H(q_2), H(p_1) - H(q_1) \}.
\label{eq:example2-9}
\end{equation}
Plugging \eqref{eq:example2-9} in \eqref{eq:example2-7} yields
\begin{equation}
\lim_{n \rightarrow \infty} \mbox{E}_{P_{X^n}}\left[ \mu\left( \delta \in [0,1) \, : \, \frac{c_n^{z|x}(\delta, \bX^n)}{n} - \frac{c_n^w(\delta, \bX^n)}{n} \leq M \right) \right] =0.
\label{eq:example2-10}
\end{equation}
\eqref{eq:example2-10} immediately implies that
\begin{equation*}
\sup\left\{\gamma \in \bar{\bbR} \, : \, \lim_{n \rightarrow \infty}\mbox{E}_{P_{X^n}}\left[ \mu\left( \delta \in [0,1) \, : \, \frac{c_n^{z|x}(\delta, \bX^n)}{n} - \frac{c_n^{w}(\delta, \bX^n)}{n} < \gamma \right)\right] = 0 \right\} \geq M,
\end{equation*}
which directly implies that
\begin{equation}
\forall \gamma \in \bbR, \, \lim_{n \rightarrow \infty}\mbox{E}_{P_{X^n}}\left[ \mu\left( \delta \in [0,1) \, : \, c_n^{z|x}(\delta, \bX^n) - c_n^{w}(\delta, \bX^n) < \gamma \right)\right] = 0.
\label{eq:example2-11}
\end{equation}
Recalling the sufficient condition for channel approximation, \eqref{eq:example2-11} implies that $\bZ$ is an approximating source for $\bW_{Y|X}$, given $\bX$.

\item \underline{$\alpha > \beta$}: Let $\bbR^+ \ni \epsilon \eqdef \frac{\alpha - \beta}{2}$. Using \eqref{eq:example2-7} and \eqref{eq:example2-8}, we have
\begin{equation}
\lim_{n \rightarrow \infty} \left\{\frac{c_n^{z|x}(\delta + \epsilon, \bX^n)}{n} - \frac{c_n^{w}(\delta, \bX^n)}{n} \right\}< - \gamma,
\label{eq:example2-12}
\end{equation}
for any $\gamma \in \left( 0,H(q_2) - H(p_1)\right)$, $\delta \in \left( \beta, \frac{\alpha + \beta}{2} \right)$ and $\bx^n$. Since $\alpha > \beta$, we have
\begin{equation}
\mu\left( \left(\beta,\frac{\alpha+\beta}{2} \right) \right) > 0.
\label{eq:example2-13}
\end{equation}
Combining \eqref{eq:example2-12} and \eqref{eq:example2-13} yields
\begin{equation}
\exists \epsilon \in (0,1), \, \exists \gamma \in \bbR^+, \, \mbox{ s.t. } \lim_{n \rightarrow \infty} \mbox{E}_{P_{X^n}} \left[ \mu\left( \delta \in [0,1 - \epsilon) \, : \, \frac{c_n^{z|x}(\delta + \epsilon, \bX^n)}{n} - \frac{c_n^{w}(\delta, \bX^n)}{n} < - \gamma \right)\right] >0.
\label{eq:example2-14}
\end{equation}
Recalling the necessary condition for channel approximation, \eqref{eq:example2-14} implies that $\bZ$ is not an approximating source for $\bW_{Y|X}$, given $\bX$.
\end{enumerate}
Note that since the channels we mix in Example 2 are memoryless binary symmetric channels, the resulting necessary and sufficient conditions are independent from the input source distribution, therefore the problem reduces to source approximation problem. Hence, Example 2 demonstrates (possibly in an exaggerated manner) the close relationship between source and channel approximation. 

\textbf{Example 3:} Let $\{\bZ_i^n\}_{n \geq 1}$ be stationary memoryless Bernoulli source with parameters $p_i$, for $i \in \{ 1,2\}$, such that $0 < p_1 < p_2 < 1/2$. For any $n \in \bbZ^+$, define 
\begin{equation}
\bZ^n \eqdef  Q_1 \bZ_1^n + (1 - Q_1)\bZ_2^n,
\label{eq:example3-1} 
\end{equation}
where $Q_1$ is a Bernoulli random variable with parameter $0 \leq \alpha \leq 1/2$. We denote the coin source as $\bZ = \{ \bZ^n\}_{n = 1}^\infty$.

Further, for any $n \in \bbZ^+$, define
\begin{equation}
\bX^n \eqdef  Q_2 \bX^n_1 + (1 - Q_2) \bX^n_2,
\label{eq:example3-2}
\end{equation}
where $Q_2$ is a Bernoulli random variable with parameter $0 < \beta \leq 1/2$ and $P_{X_i^n} = \prod_{j=1}^n P_{X_i}(x_j)$, for $i \in \{ 1,2\}$ and 
\begin{equation}
P_{X_1}(x) = \left\{ \begin{array}{cl} 1/2, & \mbox{ if }x \in \{ 0,1\},   \\
0, & \mbox{ if } x = 2,\\
\end{array} \right.
\label{eq:example3-3}
\end{equation}
and 
\begin{equation}
P_{X_2}(x) = \left\{ \begin{array}{cl} 0, & \mbox{ if } x \in \{ 0,1\},   \\
1, & \mbox{ if } x = 2.\\
\end{array} \right.
\label{eq:example3-4}
\end{equation}
We denote input source as $\bX = \{ \bX^n\}_{n=1}^\infty$. Moreover, for $i \in \{ 1, 2\}$ define the following channels 
\begin{equation}
W_{Y|X}^i(y|x) = \left\{ \begin{array}{cl} 1 - q_i, & \mbox{ if } (x,y) = (0,0) \mbox{ or } (x,y) = (1,1) , \\
q_i, & \mbox{ if } (x,y) = (0,1) \mbox{ or } (x,y) = (1,0), \\
1, & \mbox{ if } (x,y) = (2, 2) , \\
0, & \mbox{ else},
\end{array} \right.
\label{eq:example3-5}
\end{equation}
with $0 < q_1 < q_2 < 1/2$. Using \eqref{eq:example3-5}, we define the following general channel.

\begin{figure}[!htb]
\centering \centerline{\epsfxsize = 3.0in
\epsfbox{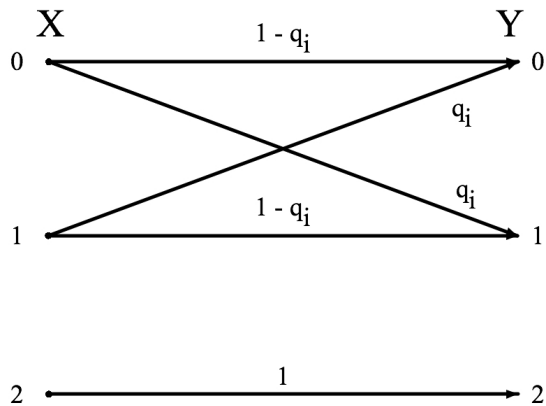}} 
\caption{Channel defined in \eqref{eq:example3-5}.} 
\label{fig:fig2}
\end{figure}

\begin{equation}
W_{Y^n|X^n} = \left\{ \begin{array}{cl} W^1_{Y^n|X^n}, & \mbox{ with probability } \alpha,   \\
W^2_{Y^n|X^n}, & \mbox{ with probability } 1 -\alpha,\\
\end{array} \right.
\label{eq:example3-6}
\end{equation}
where for $i \in \{ 1,2\}$, $W^i_{Y^n|X^n} \eqdef \prod_{j=1}^n W^i_{Y|X}(y_j|x_j)$ with $W^i_{Y|X}$ is as defined in \eqref{eq:example3-5} and $0 \leq \alpha \leq 1/2 $. Let $\bW_{Y|X} = \{ W_{Y^n|X^n}\}_{n=1}^\infty$ denote the general channel. 

Further, suppose that 
\begin{equation}
\bX_i^n \independent \bZ^n_i \independent Q_i,
\label{eq:example3-7}
\end{equation}
$\forall n \in \bbZ^+, \, \forall i \in \{ 1,2\}$. Observe that using \eqref{eq:example3-7}, we have 
\begin{equation}
c_n^{z|x}(\delta, \bx^n) = c_n^z(\delta),
\label{eq:example3-9} 
\end{equation}
for any $\delta \in [0,1)$ and $\bx^n \in \{ 0,1,2 \}^n$. Hence, \eqref{eq:example3-9} yields
\begin{equation}
\lim_{n \rightarrow \infty}\frac{c_n^{z|x}(\delta, \bx^n)}{n}  = H(p_1) \mathbf{1}_{\{ 0 < \delta < \alpha\}} + H(p_2) \mathbf{1}_{\{ \alpha < \delta < 1 \}},
\label{eq:example3-10}
\end{equation}
for any $\delta \in [0,1)$ except $\delta = \alpha$ and for any $\bx^n \in \{ 0,1,2 \}^n $. Further, (recall \eqref{eq:defn-c}) \eqref{eq:example3-4}, \eqref{eq:example3-5} and \eqref{eq:example3-6} implies that
\begin{align}
& \forall \, \delta \in [0,1)\backslash \alpha, \, \forall \, \bx^n \in \{ 0,1\}^n, \, \lim_{n \rightarrow \infty}\frac{c_n^w(\delta, \bx^n)}{n}  = H(q_1) \mathbf{1}_{\{0 < \delta < \alpha\}} + H(q_2) \mathbf{1}_{\{\alpha < \delta < 1 \}}, \label{eq:example3-12} \\
& \forall \, \delta \in [0,1), \, \forall \, \bx^n \in \{ 2\}^n, \, \lim_{n \rightarrow \infty}\frac{c_n^w(\delta, \bx^n)}{n} = 0. \label{eq:example3-13} 
\end{align}
Next, consider the following possibilities:
\begin{enumerate}
\item\underline{$H(p_2) > H(q_2)$, $H(p_1) > H(q_1)$:} Let $M \eqdef \min\{ H(p_2) - H(q_2), H(p_1) - H(q_1)\} > 0$. \eqref{eq:example3-4}, \eqref{eq:example3-12} and \eqref{eq:example3-13} yields
\begin{equation}
\lim_{n \rightarrow \infty}\mu\left( \delta \in [0,1) \, : \, \frac{c_n^{z|x}(\delta, \bX^n)}{n} - \frac{c_n^w(\delta, \bX^n)}{n} < M/2 \right) = 0 \quad \mbox{ a.s.}
\label{eq:example3-14}
\end{equation}
Moreover, since $\Pr\left\{ \mu\left( \delta \in [0,1) \, : \, \frac{c_n^{z|x}(\delta, \bX^n)}{n} - \frac{c_n^w(\delta, \bX^n)}{n} < M/2 \right) \leq 1 \right\} =1$, for all $n \in \bbZ^+$, using dominated convergence theorem, \eqref{eq:example3-14} implies that
\begin{equation*}
\lim_{n \rightarrow \infty} \mbox{E}_{P_{X^n}}\left[ \mu\left( \delta \in [0,1) \, : \, \frac{c_n^{z|x}(\delta, \bX^n)}{n} - \frac{c_n^w(\delta, \bX^n)}{n} < M/2 \right) \right] = 0, 
\end{equation*}
which in turn implies that (cf. sufficient condition for channel simulation) $\bZ$ is an approximating source for $\bW$, given $\bX$.

\item\underline{$H(p_2) < H(q_2)$ or $H(p_1) < H(q_1)$:} Let $M \eqdef \min\{ (H(q_2) - H(p_2))^+, (H(q_1) - H(p_1))^+ \} \in \bbR^+$, where $(x)^+$ denotes $\max\{x, 0\}$, for any $x \in \bbR$. Further, fix some $\epsilon \in (0, 1/4)$. Recalling the law of total expectation, we have
\begin{align}
 & \mbox{E}_{P_{X^n}}\left[ \mu\left( \delta \in [0,1-\epsilon) \, : \, \frac{c_n^{z|x}(\delta + \epsilon, \bX^n)}{n} - \frac{c_n^w(\delta, \bX^n)}{n} < -\frac{M}{2} \right) \right] = \nonumber \\
 & \sum_{i =0}^1 \Pr\{Q_2 = i\} \cdot \mbox{E}\left[ \mu\left( \delta \in [0,1-\epsilon) \, : \, \frac{c_n^{z|x}(\delta + \epsilon, \bX^n)}{n} - \frac{c_n^w(\delta, \bX^n)}{n} < -\frac{M}{2} \right) \, \big| \, Q_2 = i \right], \label{eq:example3-15}
\end{align}
for any $n \in \bbZ^+$. Further, recalling \eqref{eq:example3-10} and \eqref{eq:example3-12}, we have
\begin{equation}
\liminf_{n \rightarrow \infty} \mbox{E}\left[ \mu\left( \delta \in [0,1-\epsilon) \, : \, \frac{c_n^{z|x}(\delta + \epsilon, \bX^n)}{n} - \frac{c_n^w(\delta, \bX^n)}{n} < -\frac{M}{2} \right) \, \big|  \, Q_2 = 1 \right] >0.
\label{eq:example3-16}
\end{equation}
Plugging \eqref{eq:example3-16} into \eqref{eq:example3-15} (since $\beta >0$), we conclude that (cf. the necessary condition for source simulation) $\bX$ is \emph{not} an approximating source for $\bY$.
\item\underline{$H(p_1) = H(q_1)$, $H(p_2) = H(q_2)$:} Although it is possible to simulate $\bY$ using $\bX$, the results in this paper are inconclusive for this case.
\end{enumerate}
\textbf{Example 4:} 
Let $\cX_n = \cX$ (resp.  $\cY_n$) be some arbitrary set with $|\cX| < \infty$ (resp. $|\cY_n| = \sqrt{n}$), for all $n \in \bbZ^+$. For any $n \in \bbZ^+$, define $X_n$ (resp. $Y_n$) as the uniform random variable over $\cX_n$ (resp. $\cY_n$). Recalling \eqref{eq:defn-c}, we have
\begin{equation}
\lim_{n \rightarrow \infty} c_n^y(\delta) = \frac{1}{2}\log n, \quad \lim_{n \rightarrow \infty} c_n^x(\delta) = \log |\cX|,
\label{eq:remark-necc-cond-comp1}
\end{equation}
for all $\delta \in [0,1)$. Recalling the necessary condition for the source simulation, \eqref{eq:remark-necc-cond-comp1} immediately implies that $\{ X_n\}_{n =1}^\infty$ is \emph{not} an approximating source for $\{ Y_n \}_{n=1}^\infty$. However, if $X_n$ and $Y_n$ are independent for all $n \in \bbZ^+$, in other words if we let $P_{X_n, Y_n} = P_{X_n} P_{Y_n}$, then we have
\begin{equation}
p-\liminf_{n \rightarrow \infty} \left\{ \frac{1}{n}\log\frac{1}{P_{X_n}(X_n)} - \frac{1}{n}\log\frac{1}{P_{Y_n}(Y_n)}\right\} \geq 0.
\label{eq:remark-necc-cond-comp2}
\end{equation}
Recalling \eqref{eq:japanese-necessary}, \eqref{eq:remark-necc-cond-comp2} implies that the necessary condition of \cite{nagaoka96b} is satisfied.

\section{Source Simulation}
\label{sec:source-approximation}
In this section, we deal with the problem of approximating a general target source with a general coin source. Although this problem is a special case of the channel approximation with a deterministic input to the channel, we include it as a separate section for the following reasons: First, the results of this section constitute the core of the sufficiency and necessity proofs of channel simulation problem; to be more precise, Proposition~\ref{prop:single-source-prop-1} (resp. Theorem~\ref{thrm:thrm-source-necc}) plays a fundamental role in the proof of sufficient (resp. necessary) condition of the channel simulation problem. Moreover, for this special case, we prove a stronger necessary condition compared to the single source counterpart of the channel simulation necessary condition. Last reason is the problem's particular practical importance. 

Throughout this section, let $\bX = \{ X_n\}_{n=1}^\infty$ and $\bY = \{ Y_n \}_{n=1}^\infty$ be two general sources with $\cX_n$ and $\cY_n$ being countable sets for all $n \in \bbZ^+$.

\subsection{Source Simulation-Sufficient Condition}
\label{ssec:source-approximation-suff-cond}
First, we begin with the following proposition, which constitutes the core of the achievability proofs for both source and channel approximation problems.
\begin{proposition}
Consider any $\epsilon \in (0,1)$ and $\gamma \in \bbR^+$, such that $e^{- \gamma} \leq \epsilon$. Then, for any $n \in \bbZ^+$,  
\[
\exists \, \phi_n: \cX_n \rightarrow \cY_n, \mbox{ with } d(Y_n, \phi_n(X_n)) \leq 9 \epsilon + 10 \mu(\cE^n(\gamma)),
\]
where $\cE^n(\gamma) \eqdef \{ \delta \in [0,1) \, : \, c_n^x(\delta) - c_n^y(\delta) < \gamma\}$.
\label{prop:single-source-prop-1}
\end{proposition}

\begin{proof}
Consider any $\epsilon \in (0,1)$ and $\gamma \in \bbR^+$, such that $e^{- \gamma} \leq \epsilon$ and consider any $n \in \bbZ^+$. Define $\cA^n(\gamma) \eqdef [0,1) \backslash \cE^n(\gamma)$. Next, fix some $\cS_n(X)$ and $\cS_n(Y)$ as given in Definition~\ref{def:c-n} and define the following indexes:
\begin{align}
i_1 \eqdef \inf\{ j \in \bbZ^+ \, : \, \sum_{i = j+1}^\infty P_{X_n}(x_i) < \epsilon \}, \label{eq:prop-1-proof2} \\
i_2 \eqdef \inf\{ j \in \bbZ^+ \, : \, \sum_{i = j+1}^\infty P_{Y_n}(y_i) < \epsilon \}. \label{eq:prop-1-proof3} 
\end{align}
Note that both $i_1 < \infty $ and $i_2 < \infty$. 

Define $\eta_x = \sum_{i = i_1 +1}^{\infty} P_{X_n}(x_i)$ (resp. $\eta_y = \sum_{i = i_2 +1}^{\infty} P_{Y_n}(y_i) $). Using these, define the following partitions of $[0,1]$
\begin{align}
\Delta^x & \eqdef \{ 0 = \delta_0^x < \delta_1^x < \ldots < \delta_{i_1}^x = 1 - \eta_x \}, \, \mbox{s.t. } \, \forall i, \Delta_i^x = \delta_i^x - \delta_{i-1}^x = P_{X_n}(x_i), \label{eq:prop-1-proof4} \\
\Delta^y & \eqdef \{ 0 = \delta_0^y < \delta_1^y < \ldots < \delta_{i_2}^y = 1 - \eta_y \}, \, \mbox{s.t. } \, \forall i, \Delta_i^y = \delta_i^y - \delta_{i-1}^y = P_{Y_n}(y_i). \label{eq:prop-1-proof5}
\end{align}

Next, we define the following mapping:
\begin{equation}
\phi_n^\prime \, : \, \{ x_i \}_{i=1}^{i_1} \rightarrow \{ y_i \}_{i=1}^{i_2}, \, \mbox{s.t. } \, \phi_n^\prime(x_i) = y_j, \mbox{ if } \delta_{i-1}^x \in [ \delta_{j-1}^y, \delta_j^y), \, \forall i \in \{ 1, \ldots, i_1 \},
\label{eq:prop-1-proof6}
\end{equation} 
for some corresponding $j \in \{ 1, \ldots, i_2\}$. We define $j_2 \in \{1, \ldots, i_2 \}$ as the only index with $\phi_n^\prime(x_{i_1}) = y_{j_2}$. Note that $j_2 \leq i_2$. It can easily be verified that $\phi_n^\prime$ is a well-defined mapping. Now, using \eqref{eq:prop-1-proof6}, we define $\phi_n \, : \, \cX_n \rightarrow \cY_n $, such that
\begin{equation}
\phi\left( x \right) = \left\{ \begin{array}{cl}\phi^\prime_n(x_i), & \mbox{ if } x \in \{ x_i\}_{i=1}^{i_1}  \\
y_{i_2}, & \mbox{ if } x \notin \{ x_i\}_{i=1}^{i_1} \\
\end{array} \right.
\label{eq:prop-1-proof7}
\end{equation}
As a shorthand, let $\tilde{Y}_n \eqdef \phi_n(X_n)$ denote the output of the mapping. 

After defining the mapping, we analyze the variational distance between $Y_n$ and $\tY_n$. First of all, using \eqref{eq:prop-1-proof7}, we have
\begin{equation}
d(Y_n, \tY_n) \leq \sum_{i=1}^{j_2} |P_{Y_n}(y_i) - P_{\tY_n}(y_i)| + |P_{Y_n}(y_{i_2}) - P_{\tY_n}(y_{i_2})| + \sum_{i = j_2+1}^\infty P_{Y_n}(y_i).
\label{eq:prop-1-proof8}
\end{equation}

Next, we need the following sets of indices:
\begin{align*}
\cI & \eqdef \{ i \in \{ 1, \ldots, j_2\}\, : \, P_{\tY_n}(y_i) \neq 0 \}, \\
\tilde{\cI} & \eqdef \{ i \in \{ 1, \ldots, i_2\}\, : \, \Pr( (\phi_n^\prime)^{-1}(y_i)  ) = 0 \}.
\end{align*}

Further, we define the following subsets of $\cE^n(\gamma)$:
\begin{align}
& \forall i \in \{ 1, \ldots, j_2 -1 \}, \, \cE_{i} \eqdef (\delta_i^y, \delta_k^x)\cap \cE^n(\gamma), \mbox{ with } \delta_i^y \in [\delta_{k-1}^x, \delta_{k}^x) \mbox{ for some } k \in \bbZ^+,\label{eq:prop-1-proof9}\\
& \cE_{j_2} \eqdef(\delta_{j_2}^y, 1 - \epsilon)\cap \cE^n(\gamma). \label{eq:prop-1-proof10}\\
& \forall i \in \{1, \ldots, i_2-1 \}, \, \tilde{\cE}_i \eqdef (\delta_{i-1}^y, \delta_i^y)\cap \cE^n(\gamma), \label{eq:prop-1-proof11} \\
& \tilde{\cE}_{i_2} \eqdef (\delta_{i_2-1}^y, 1 - \epsilon) \cap \cE^n(\gamma). \label{eq:prop-1-proof12} 
\end{align}

\begin{remark}
Note that, in some sense, the aforementioned sets can be thought of as the ``bad sets'', which will contribute to the variational distance. Hence, we aim to relate the resulting variational distance between $Y_n$ and $\tY_n$ created by \eqref{eq:prop-1-proof7} to the measure of these sets.  Observe that $\forall i,j \in \{ 1, \ldots, i_2 \}, \, \tilde{\cE}_i \cap \tilde{\cE}_j = \emptyset$, if $i \neq j$, i.e. $\{ \tilde{\cE}_i \}_{i=1}^{i_2}$ is a disjoint collection of subsets belonging to $\cE^n(\gamma)$.
\end{remark}

In order to upper-bound \eqref{eq:prop-1-proof8}, we first prove the following lemmata:
\begin{lemma}
For any $i \in \{1, \ldots, j_2-1\} $, if $\delta_i^y \in (\delta_{k-1}^x, \delta_k^x)$ for some $k \in \bbZ^+$, then 
\begin{equation}
\delta_k^x - \delta_i^y \leq \epsilon P_{Y_n}(y_{i+1}) + \mu(\cE_i).
\label{eq:lem1}
\end{equation}
\label{lem:lem1}
\end{lemma}
\vspace{-0.5cm}
\begin{proof}

Consider any $i \in \{ 1, \ldots, j_2-1 \}$ and suppose $\exists \, \delta \in (\delta_{i}^y, \delta_{k}^x )$, such that $\delta \notin \cE_{i}$, i.e. $\delta \in \cA^n(\gamma)$, where $\cE_{i}$ is as defined in \eqref{eq:prop-1-proof9}. Then, recalling \eqref{eq:defn-c} and the ordering provided by $\cS_n(Y)$, we have
\begin{align}
c_n^x(\delta) = & \log \frac{1}{P_{X_n}(x_{k})}, \label{eq:lem1-proof-1}\\
c_n^y( \delta) \geq & \log \frac{1}{P_{Y_n}(y_{i+1})}. \label{eq:lem1-proof-2}
\end{align}
Using definition of $\cA^n(\gamma)$, we have
\begin{equation}
P_{X_n}(x_{k}) \leq e^{-\gamma} P_{Y_n}(y_{i+1}) \leq \epsilon P_{Y_n}(y_{i+1}),
\label{eq:lem1-proof-3}
\end{equation}
Further, note that $P_{X_n}(x_{k}) \geq \delta_{k}^x - \delta_{i}^y$. Combining this with \eqref{eq:lem1-proof-3} yields, 
\begin{equation}
\delta_{k}^x - \delta_{i}^y \leq \epsilon P_{Y_n}(y_{i+1}).
\label{eq:lem1-proof-4}
\end{equation}

If $\nexists \, \delta \in ( \delta_{i}^y, \delta_{k}^x)$, such that $\delta \notin \cE_{i}$, then we have $(\delta_{i}^y, \delta_{k}^x) = \cE_{i}$, which readily implies
\begin{equation}
\delta_{k}^x - \delta_{i}^y = \mu(\cE_{i}).
\label{eq:lem1-proof-5}
\end{equation}

Combining \eqref{eq:lem1-proof-4} and \eqref{eq:lem1-proof-5} yields
\begin{equation*}
\delta_{k}^x - \delta_{i}^y \leq \epsilon P_{Y_n}(y_{i+1})  + \mu(\cE_{i}),
\end{equation*}
which was to be shown. 
\end{proof}

\begin{lemma}
For any $i \in \tilde{\cI}$ such that $i \neq i_2$ 
\begin{equation}
P_{Y_n}(y_i) = \mu(\tilde{\cE}_i).
\label{eq:lem2}
\end{equation}
Further, if $i_2 \in \tilde{\cI}$, then
\begin{equation}
P_{Y_n}(y_{i_2}) \leq 2 \epsilon + \mu(\tilde{\cE}_{i_2}).
\label{eq:lem2-2}
\end{equation}
\label{lem:lem2}
\end{lemma}
\begin{proof}

Consider any $i \in \tilde{\cI}$. 

First, suppose $i \neq i_2$ and observe that by recalling \eqref{eq:prop-1-proof6}, $i \in \tilde{\cI}$ implies that
\begin{equation}
[\delta_{i-1}^y, \delta_i^y) \subset (\delta_{k-1}^x, \delta_k^x),
\label{eq:lem2-proof-1}
\end{equation}
for some $k \in \bbZ^+$. Next, suppose that $\exists \delta \in (\delta_{i-1}^y, \delta_i^y)$, such that $\delta \notin \tilde{\cE}_i$, i.e. $\delta \in \cA^n(\gamma)$. Then, recalling \eqref{eq:defn-c} we have
\begin{align}
c_n^x( \delta) = & \log \frac{1}{P_{X_n}(x_{k})}, \label{eq:lem2-proof-2}\\
c_n^y(\delta) = & \log \frac{1}{P_{Y_n}(y_{i})}. \label{eq:lem2-proof-3}
\end{align}
Hence, \eqref{eq:lem2-proof-2} and \eqref{eq:lem2-proof-3} implies that we have
\begin{equation*}
\delta_k^x - \delta_{k-1}^x = P_{X_n}(x_k) \leq e^{-\gamma} P_{Y_n}(y_i) \leq \epsilon (\delta_i^y - \delta_{i-1}^y) <  \delta_i^y - \delta_{i-1}^y,
\end{equation*}
which yields a contradiction with \eqref{eq:lem2-proof-1}, therefore we conclude that 
\begin{equation}
(\delta_{i-1}^y , \delta_i^y) = \tilde{\cE}_i.
\label{eq:lem2-proof-4}
\end{equation}
\eqref{eq:lem2-proof-4} directly implies that $\delta_{i}^y - \delta_{i-1}^y = P_Y(y_i) = \mu(\tilde{\cE}_i)$, which proves \eqref{eq:lem2}.

Next, suppose $i_2 \in \tilde{\cI}$, which directly implies 
\begin{align}
\delta_{i_1-1}^x  & < \delta_{i_2-1}^y, \label{eq:lem2-proof-5} \\
P_{Y_n}(y_{i_2}) & = (\delta_{i_2}^y - 1 + \epsilon) + (1 - \epsilon - \delta_{i_2-1}^y), \nonumber \\
 & \leq \epsilon + (1 - \epsilon - \delta_{i_2-1}^y), \label{eq:lem2-proof-6}
\end{align} 
where \eqref{eq:lem2-proof-6} follows from \eqref{eq:prop-1-proof3}. 

Next, suppose $\exists \delta \in ( \delta_{i_2-1}^y, 1 - \epsilon)$, such that $\delta \notin \tilde{\cE}_{i_2}$, where $\tilde{\cE}_{i_2}$ is as defined in \eqref{eq:prop-1-proof12}. Then, recalling \eqref{eq:defn-c} and \eqref{eq:lem2-proof-5} we have
\begin{align}
c_n^x(\delta) = & \log \frac{1}{P_{X_n}(x_{i_1})}, \label{eq:lem2-proof-7}\\
c_n^y(\delta) = & \log \frac{1}{P_{Y_n}(y_{i_2})}. \label{eq:lem2-proof-8}
\end{align}
Hence, \eqref{eq:lem2-proof-7} and \eqref{eq:lem2-proof-8} implies that we have
\begin{equation}
P_{X_n}(x_{i_1}) \leq e^{-n \gamma} P_{Y_n}(y_{i_2}) \leq \epsilon P_{Y_n}(y_{i_2}) \leq \epsilon.
\label{eq:lem2-proof-9}
\end{equation}
Moreover,
\begin{equation}
P_{X_n}(x_{i_1}) > (1 - \epsilon - \delta_{i_2-1}^y),
\label{eq:lem2-proof-10} 
\end{equation}
which follows from \eqref{eq:lem2-proof-5}. 
Combining \eqref{eq:lem2-proof-9} and \eqref{eq:lem2-proof-10} yields
\begin{equation}
(1 - \epsilon - \delta_{i_2-1}^y) \leq \epsilon.
\label{eq:lem2-proof-11}
\end{equation}

If $\nexists \delta \in ( \delta_{i_2-1}^y, 1 - \epsilon)$, such that $\delta \notin \tilde{\cE}_{i_2}$, then we have (cf. \eqref{eq:prop-1-proof12})
\begin{equation}
1- \epsilon - \delta_{i_2-1}^y = \mu(\tilde{\cE}_{i_2}).
\label{eq:lem2-proof-12}
\end{equation}

\eqref{eq:lem2-proof-11} and \eqref{eq:lem2-proof-12} implies that we have 
\begin{equation}
1- \epsilon - \delta_{i_2-1}^y \leq \epsilon + \mu(\tilde{\cE}_{i_2}).
\label{eq:lem2-proof-13}
\end{equation}

Plugging \eqref{eq:lem2-proof-13} into \eqref{eq:lem2-proof-6} yields 
\begin{equation*}
P_{Y_n}(y_{i_2})  \leq 2 \epsilon + \mu(\tilde{\cE}_{i_2}),
\end{equation*}
which was to be shown.
\end{proof}

Next, in order to rewrite $d(Y_n, \tY_n)$ in terms of $\cE^n(\gamma)$, we need disjointness relations between elements of $\{ \cE_i\}_{i \in \cI}$, even if this collection may not be necessarily pairwise disjoint. To this end, we need following sets:
\begin{eqnarray}
\cI_{0,0} & \eqdef & \{ i \in \cI \, : \, 1 \leq i \leq j_2-1, \, \exists k_1 \in \bbZ^+ \mbox{ s.t. }\delta_{i-1}^y \in (\delta_{k_1-1}^x, \delta_{k_1}^x), \, \exists k_2 \in \bbZ^+ \mbox{ s.t. } \delta_{i}^y \in (\delta_{k_2-1}^x, \delta_{k_2}^x)  \} \label{eq:suffi-proof-I1}\\
\cI_{0,1} & \eqdef & \{ i \in \cI \, : \, 1 \leq i \leq j_2-1, \, \exists k_1 \in \bbZ^+ \mbox{ s.t. } \delta_{i-1}^y \in (\delta_{k_1-1}^x, \delta_{k_1}^x), \, \exists k_2 \in \bbZ^+ \mbox{ s.t. } \delta_{i}^y = \delta_{k_2}^x \} \label{eq:suffi-proof-I2}\\
\cI_{1,0} & \eqdef & \{ i \in \cI \, : \, 1 \leq i \leq j_2-1, \, \exists k_1 \in \bbZ^+ \mbox{ s.t. } \delta_{i-1}^y = \delta_{k_1}^x, \, \exists k_2 \in \bbZ^+ \mbox{ s.t. } \delta_{i}^y \in (\delta_{k_2-1}^x, \delta_{k_2}^x)  \} \label{eq:suffi-proof-I3}\\
\cI_{1,1} & \eqdef & \{ i \in \cI \, : \, 1 \leq i \leq j_2-1, \, \exists k_1 \in \bbZ^+ \mbox{ s.t. } \delta_{i-1}^y = \delta_{k_1}^x, \, \exists k_2 \in \bbZ^+ \mbox{ s.t. } \delta_{i}^y = \delta_{k_2}^x  \} \label{eq:suffi-proof-I4}
\end{eqnarray}
Note that $\{ \cI_i\}_{i=1}^4$ forms a partition of $\cI \backslash \{ j_2\}$.

First, suppose $i \in \{ 1, \ldots, j_2-1\}$. 

Now, if $i \in \cI \backslash j_2$, then using \eqref{eq:prop-1-proof6}, we have
\begin{equation}
| P_{Y_n}(y_i) - P_{\tY_n}(y_i)| \leq \left\{ \begin{array}{cl}(\delta_{k_1}^x - \delta_{i-1}^y ) + (\delta_{k_2}^x - \delta_{i}^y), & \mbox{ if } i \in \cI_{0,0},  \\
(\delta_{k_1}^x - \delta_{i-1}^y ) , & \mbox{ if } i \in \cI_{0,1} \\
(\delta_{k_2}^x - \delta_{i}^y ) , & \mbox{ if } i \in \cI_{1,0}, \\
0, & \mbox{ if } i \in \cI_{1,1}. \\
\end{array} \right.
\label{eq:prop-1-proof17}
\end{equation}

We have
\begin{align}
\sum_{i \in \cI \backslash j_2}|P_{Y_n}(y_i) - P_{\tY_n}(y_i)| & = \sum_{k=1}^3\sum_{i \in \cI_k}|P_{Y_n}(y_i) - P_{\tY_n}(y_i)| , \label{eq:prop-1-proof18} \\
 & \leq \sum_{i \in \cI_{0,0}} \left( \epsilon (P_{Y_n}(y_i) + P_{Y_n}(y_{i+1}) ) + \mu(\cE_i) + \mu(\cE_{i-1}) \right) + \sum_{i \in \cI_{0,1}}\left( \epsilon P_{Y_n}(y_i) + \mu(\cE_{i-1}) \right)\nonumber \\ 
 & + \sum_{i \in \cI_{1,0}} \left( \epsilon P_{Y_n}(y_{i+1}) + \mu(\cE_i)\right), \label{eq:prop-1-proof19} \\
& \leq 2 \epsilon \sum_{i=1}^{j_2-1} P_{Y_n} (y_i) + \sum_{i \in \cI_{0,0}}( \mu(\cE_i) + \mu(\cE_{i-1}) )+ \sum_{i \in \cI_{0,1}} \mu(\cE_{i-1}) + \sum_{i \in \cI_{1,0}} \mu(\cE_i), \label{eq:prop-1-proof20}
\end{align}
where \eqref{eq:prop-1-proof18} follows from \eqref{eq:prop-1-proof17}, \eqref{eq:prop-1-proof19} follows from \eqref{eq:lem1}, \eqref{eq:prop-1-proof20} follows from definition of $\cS_n(Y)$ and disjointness of $\cI_{0,0}, \cI_{0,1}, \cI_{1,0}$. Observe that $1 \in \cI_{1,0} \cup \cI_{1,1}$ from the definition, so in \eqref{eq:prop-1-proof20} we never refer to $\mu(\cE_0)$, which is not defined. 

In order to bound \eqref{eq:prop-1-proof20}, we need following result.
\begin{lemma}
$\{ \cE_ i\}_{i \in \cI_{0,0}}$, $\{ \cE_{i-1} \}_{i \in \cI_{0,0}}$, $\{ \cE_{i-1}\}_{i \in \cI_{0,1}}$ and $\{ \cE_i \}_{i \in \cI_{1,0}}$ are disjoint collection of subsets of $\cE^n(\gamma)$.
\label{lem:lem3}
\end{lemma}
 
\begin{proof}

We prove following two claims, which are sufficient to conclude the result.
\begin{claim}
For any $i, j \in \cI$, such that $i, j \neq j_2$ and $i \neq j$, if $\exists k \in \bbZ^+$ with $\delta_i^y \in (\delta_{k-1}^x, \delta_k^x)$ and $\exists l \in \bbZ^+$ with $\delta_j^y \in (\delta_{l-1}^x, \delta_l^x)$, then $\cE_i \cap \cE_j = \emptyset$.
\label{cla:cla1}
\end{claim}

\begin{proof}

Consider any $i, j \in \cI$, such that $i, j \neq j_2$ and $i \neq j$, with $\delta_i^y \in (\delta_{k-1}^x, \delta_k^x)$ and $\delta_j^y \in (\delta_{l-1}^x, \delta_l^x)$, for some $k, l \in \bbZ^+$. W.l.o.g. suppose $i < j$, which implies that $k \leq l$. If $k = l$, we have $[\delta_{j-1}^y, \delta_j^y] \subset (\delta_{k-1}^x, \delta_k^x)$, which contradicts $j \in \cI$, hence we should have $k < l$, which immediately implies that $(\delta_i^y, \delta_k^x) \cap (\delta_j^y, \delta_l^x) = \emptyset$, which was to be shown.
\end{proof}

\begin{claim}
For any $i, j \in \cI$, such that $i, j \neq j_2$ and $i \neq j$, if $\exists k \in \bbZ^+$ with $\delta_{i-1}^y \in (\delta_{k-1}^x, \delta_k^x)$ and $\exists l \in \bbZ^+$ with $\delta_{j-1}^y \in (\delta_{l-1}^x, \delta_l^x)$, then $\cE_{i-1} \cap \cE_{j -1} = \emptyset$.
\label{cla:cla2}
\end{claim}

\begin{proof}

Consider any $i, j \in \cI$, such that $i, j \neq j_2$ and $i \neq j$, with $\delta_{i-1}^y \in (\delta_{k-1}^x, \delta_k^x)$ and $\delta_{j-1}^y \in (\delta_{l-1}^x, \delta_l^x)$, for some $k, l \in \bbZ^+$. W.l.o.g. suppose $i < j$, which immediately implies that $k \leq l$. Suppose, for contradiction, $k = l$. Since $i < j$, we have $\delta_i^y \leq \delta_{j-1}^y$, which implies that $[\delta_{i-1}^y, \delta_i^y ] \subset (\delta_{k-1}^x, \delta_k^x)$, which in turn implies that $i \in \tilde{\cI}$, which contradicts the assumption $i \in \cI$, hence we have $k < l$. Using this, we have $\delta_k^x \leq \delta_{l-1}^x$, which implies that $(\delta_{i-1}^y, \delta_k^x) \cap (\delta_{j-1}^y, \delta_l^x) = \emptyset$, which implies the result by recalling \eqref{eq:prop-1-proof9}. 
\end{proof}

Now, observe that Claim~\ref{cla:cla1} implies that $\{ \cE_i \}_{i \in \cI_{0,0}}, \{ \cE_i \}_{i \in \cI_{1,0}}$ are disjoint collection of sets, while Claim~\ref{cla:cla2} implies that $\{ \cE_{i-1}\}_{i \in \cI_{0,0}}, \{ \cE_{i-1}\}_{i \in \cI_{0,1}}$ are disjoint collection of sets, which was to be shown. 
\end{proof}

Using the result of Lemma~\ref{lem:lem3} in \eqref{eq:prop-1-proof20} we have
\begin{equation}
\sum_{\cI \backslash \{ j_2 \}} |P_{Y_n}(y_i) - P_{\tY_n}(y_i)| \leq 2 \epsilon \sum_{i=1}^{j_2-1}P_{Y_n}(y_i) + 4 \mu(\cE^n(\gamma)).
\label{eq:prop-1-proof-D-1}
\end{equation}
 
Next, suppose $i \in \{ 1, \ldots, j_2-1\}$ and $i \in \tilde{\cI}$. We have (note that $1 \in \cI$ as a result of \eqref{eq:prop-1-proof6})
\begin{align}
\sum_{1< i < j_2, \mbox{ s.t. } i \in \tilde{\cI}} |P_{Y_n}(y_i) - P_{\tY_n}(y_i)| & = \sum_{1< i < j_2, \mbox{ s.t. } i \in \tilde{\cI}}P_{Y_n}(y_i), \label{eq:prop-1-proof-D-2-temp1} \\
 & \leq \sum_{1< i < j_2, \mbox{ s.t. } i \in \tilde{\cI}} \mu(\tilde{\cE}_i), \label{eq:prop-1-proof-D-2-temp2}\\
 & \leq \mu(\cE^n(\gamma)), \label{eq:prop-1-proof-D-2}
\end{align}
where \eqref{eq:prop-1-proof-D-2-temp1} follows from definition of $\tilde{\cI}$, \eqref{eq:prop-1-proof-D-2-temp2} follows from Lemma~\ref{lem:lem2} and \eqref{eq:prop-1-proof-D-2} follows since $\{ \tilde{\cE}_i \}_{i=1}^{i_2}$ is a disjoint collection of subsets of $\cE^n(\gamma)$.

Now, we complete the proof. First, suppose that $j_2 = i_2$. Using \eqref{eq:prop-1-proof7}, we have
\begin{equation}
|P_{Y_n}(y_{i_2}) - P_{\tY_n}(y_{i_2})| \leq | \Pr\{ (\phi_n^\prime)^{-1}(y_{i_2}) \} - P_{Y_n}(y_{i_2})  | + \sum_{i=i_1+1}^\infty P_{X_n}(x_i).
\label{eq:prop-1-proof21}
\end{equation}
Further, using the result of Lemma~\ref{lem:lem1} and $|\delta_{i_2}^y - \delta_{i_1}^x| \leq \epsilon$ (cf. definition of $i_1$ and $i_2$) we have,  
\begin{equation}
| P_{Y_n}(y_{i_2}) - \Pr\{ (\phi_n^\prime)^{-1}(y_{i_2}) \}| \leq \epsilon P_{Y_n}(y_{i_2}) +\epsilon + \mu(\cE^n(\gamma)).
\label{eq:prop-1-proof23}
\end{equation}
Moreover, the definition of $i_1$ immediately implies that 
\begin{equation}
\sum_{i=i_1+1}^\infty P_{X_n}(X_n) \leq \epsilon.
\label{eq:prop-1-proof-D-3}
\end{equation} 
Plugging \eqref{eq:prop-1-proof-D-3} and \eqref{eq:prop-1-proof23} into \eqref{eq:prop-1-proof21} yields
\begin{equation}
|P_{Y_n}(y_{i_2}) - P_{\tY_n}(y_{i_2})| \leq \epsilon P_{Y_n}(y_{i_2}) + 3 \epsilon + \mu(\cE^n(\gamma)).
\label{eq:prop-1-proof24}
\end{equation}
Further, recalling definition of $i_2$, we have
\begin{equation}
\sum_{i = i_2 + 1}^\infty P_{Y_n}(y_i) \leq \epsilon.
\label{eq:prop-1-proof-D-3.5}
\end{equation}

Plugging \eqref{eq:prop-1-proof-D-1}, \eqref{eq:prop-1-proof-D-2}, \eqref{eq:prop-1-proof24} and \eqref{eq:prop-1-proof-D-3.5} into \eqref{eq:prop-1-proof8} yields
\begin{equation}
d(Y_n, \tY_n) \leq 9 \epsilon + 7 \mu(\cE^n(\gamma)), \label{eq:prop-1-proof-final1}
\end{equation}
 
Next, suppose that $j_2 < i_2$. First of all, note that $\delta_{j_2}^y = \delta_k^x$ is not possible for any $k \in \{ 1, \ldots, i_1-1\}$, because if it is the case, then $\phi_n^\prime(x_k) = y_{j_2+1}$, which contradicts with the definition of $j_2$. Further, observe that $\phi_n(x_{i_1}) = y_{j_2}$, which directly follows from \eqref{eq:prop-1-proof6}. Using these observations, we have
\begin{equation}
| P_{Y_n}(y_{j_2}) - P_{\tY_n}(y_{j_2})| = \left\{ \begin{array}{cl}(\delta_{k}^x - \delta_{j_2-1}^y ) + ( \delta_{i_1}^x - \delta_{j_2}^y) , & \mbox{ if } \delta_{j_2-1}^y \in (\delta_{k-1}^x, \delta_{k}^x),  \\
(\delta_{i_1}^x - \delta_{j_2}^y), & \mbox{ if } \delta_{j_2-1}^y = \delta_{k-1}^x,\\
\end{array} \right.
\label{eq:prop-1-proof25}
\end{equation}
for some $k \leq i_1$ with $\delta_{i_2-1}^y \in [\delta_{k-1}^x, \delta_k^x)$.

If $\exists \, \delta \in (\delta_{j_2}^y , 1 - \epsilon) $, such that $\delta \notin \cE_{j_2}$, then using similar arguments as in the proof of Lemma~\ref{lem:lem1}, it can be shown that
\begin{equation}
1 - \epsilon - \delta_{j_2}^y \leq \epsilon P_{Y_n}(y_{j_2+1}).
\label{eq:prop-1-proof26}
\end{equation} 
If $\nexists \, \delta \in (\delta_{j_2}^y , 1 - \epsilon)$, such that $ \delta \notin \cE_{j_2}$, then $( \delta_{j_2}^y, 1 - \epsilon) \subseteq \cE_{j_2}$, which implies that $ 1 - \epsilon - \delta_{j_2}^y \leq \mu(\cE_{j_2}) \leq \mu(\cE^n(\gamma))$. Combining this with \eqref{eq:prop-1-proof26} yields,
\begin{equation}
1 - \epsilon - \delta_{j_2}^y \leq \epsilon P_{Y_n}(y_{j_2+1}) +\mu(\cE^n(\gamma)).
\label{eq:prop-1-proof27}
\end{equation}
Further, $\delta_{i_1}^x - 1 + \epsilon \leq \epsilon$. Combining this with \eqref{eq:prop-1-proof27} and noting that $P_{Y_n}(y_{j_2+1}) \leq P_{Y_n}(y_{j_2}) $ (cf. definition of $\cS_n(Y)$) yields, 
\begin{equation}
\delta_{i_1}^x - \delta_{j_2}^y \leq \epsilon + \epsilon P_{Y_n}(y_{j_2}) + \mu(\cE^n(\gamma)).
\label{eq:prop-1-proof28}
\end{equation}
Using \eqref{eq:prop-1-proof28} in \eqref{eq:prop-1-proof25} and recalling the result of Lemma~\ref{lem:lem1} yields
\begin{equation}
|P_{Y_n}(y_{j_2}) - P_{\tY_n}(y_{j_2})| \leq 2 \epsilon P_{Y_n}(y_{j_2}) + \epsilon + 2\mu(\cE^n(\gamma)).
\label{eq:prop-1-proof-D-4}
\end{equation}

Next, using \eqref{eq:prop-1-proof7} and recalling the assumption of $j_2 < i_2$, we have
\begin{equation}
|P_{Y_n}(y_{i_2}) - P_{\tY_n}(y_{i_2})| \leq P_{Y_n}(y_{i_2}) + (1 - \delta_{i_1}^x) \leq 3\epsilon + \mu(\cE^n(\gamma)),
\label{eq:prop-1-proof-D-5}
\end{equation}
where second inequality follows from Lemma~\ref{lem:lem2} and the definition of $i_1$. 

Further, 
\begin{align}
\sum_{i = j_2 +1}^\infty P_{Y_n}(y_i) & = \sum_{i = j_2 + 1}^{i_2} P_{Y_n}(y_i) + \sum_{i = i_2 + 1}^\infty P_{Y_n}(y_i) \nonumber \\  
 & \leq \sum_{i = j_2 + 1}^{i_2} P_{Y_n}(y_i) + \epsilon, \label{eq:prop-1-proof29}
\end{align}
where \eqref{eq:prop-1-proof29} follows from \eqref{eq:prop-1-proof-D-3.5}.

Now, suppose $j_2 = i_2 -1$. Then, \eqref{eq:prop-1-proof29} reduces to 
\begin{equation}
\sum_{i = j_2 +1}^\infty P_{Y_n}(y_i) \leq P_{Y_n}(y_{i_2}) + \epsilon \leq 3 \epsilon + \mu(\cE^n(\gamma)),
\label{eq:prop-1-proof-D-6}
\end{equation}  
where second inequality follows using Lemma~\ref{lem:lem2}. Hence, plugging \eqref{eq:prop-1-proof-D-1}, \eqref{eq:prop-1-proof-D-2}, \eqref{eq:prop-1-proof-D-4}, \eqref{eq:prop-1-proof-D-5} and \eqref{eq:prop-1-proof-D-6} into \eqref{eq:prop-1-proof8} yields,
\begin{equation}
d(Y_n, \tY_n) \leq 9 \epsilon + 9 \mu(\cE^n(\gamma)).
\label{eq:prop-1-proof-final2}
\end{equation}

Finally, suppose $j_2 < i_2 -1$, using Lemma~\ref{lem:lem2} in \eqref{eq:prop-1-proof29}, we have
\begin{align}
\sum_{i = j_2 +1}^\infty P_{Y_n}(y_i) & \leq \sum_{i = j_2 +1}^{i_2-1} \mu(\tilde{\cE}_i) + 3 \epsilon + \mu(\cE^n(\gamma)), \nonumber \\
 & \leq 3 \epsilon + 2 \mu(\cE^n(\gamma)), \label{eq:prop-1-proof-D-7}
\end{align}
where \eqref{eq:prop-1-proof-D-7} follows from the fact that $\{ \tilde{\cE}_i \}_{i = 1}^{i_2}$ is a disjoint collection of subsets of $\cE^n(\gamma)$.  Hence, plugging \eqref{eq:prop-1-proof-D-1}, \eqref{eq:prop-1-proof-D-2}, \eqref{eq:prop-1-proof-D-4}, \eqref{eq:prop-1-proof-D-5} and \eqref{eq:prop-1-proof-D-7} into \eqref{eq:prop-1-proof8} yields,
\begin{equation}
d(Y_n, \tY_n) \leq 9 \epsilon + 10 \mu(\cE^n(\gamma)).
\label{eq:prop-1-proof-final3}
\end{equation}

Lastly, combining \eqref{eq:prop-1-proof-final1}, \eqref{eq:prop-1-proof-final2} and \eqref{eq:prop-1-proof-final3} yields the sought after result.
\end{proof}


\begin{theorem}
If 
\begin{equation}
\mu-\liminf_{n \rightarrow \infty} \{ c_n^x(\delta) - c_n^y(\delta) \} = \infty
\label{eq:thrm-suffi-source}
\end{equation}
then $\bX$ is an approximating source for $\bY$.
\label{thrm:thrm-suffi-source}
\end{theorem}

\begin{proof}

First, observe that \eqref{eq:thrm-suffi-source} is equivalent to
\begin{equation}
\forall \gamma \in \bbR, \forall \epsilon \in \bbR^+, \, \exists \, N(\epsilon, \gamma) \in \bbZ^+, \mbox{ s.t. } \, \forall n \geq N(\epsilon, \gamma), \mu \left( \delta \in [0,1) \, : \, c_n^x(\delta) - c_n^y(\delta) < \gamma \right) \leq \epsilon.
\label{eq:thrm-suffi-source-proof-1} 
\end{equation}
Next, consider any $\epsilon \in (0,1)$ and $\gamma \in \bbR^+$, such that $e^{- \gamma} \leq \epsilon$ and fix some $n \geq N(\epsilon, \gamma)$. \eqref{eq:thrm-suffi-source-proof-1} immediately implies that (recall definition of $\cE^n(\gamma)$ in Proposition~\ref{prop:single-source-prop-1})
\begin{equation}
\mu(\cE^n(\gamma)) \leq \epsilon.
\label{eq:thrm-suffi-source-proof-2}
\end{equation}
Proposition~\ref{prop:single-source-prop-1} immediately implies that
\begin{equation}
\exists \, \phi_n : \cX_n \rightarrow \cY_n, \mbox{ with } d(Y_n, \tY_n) \leq 19 \epsilon,
\label{eq:thrm-suffi-source-proof-3}
\end{equation}
where $\tY_n = \phi_n(Y_n)$. 
Next, we define
\begin{equation}
\forall n \in \bbZ^+, \, d_n^\ast \eqdef \inf\{ d(Y_n, \tY_n) \, : \, \exists \, \phi_n : \cX_n \rightarrow \cY_n\}.
\label{eq:thrm-suffi-source-proof-4}
\end{equation}
\eqref{eq:thrm-suffi-source-proof-3} implies that $\forall n \geq N(\epsilon, \gamma)$, $d^\ast_n \leq 19 \epsilon$. Since $\epsilon \in (0,1)$ is arbitrary, this implies that we have
\begin{equation}
\limsup_{n \rightarrow \infty} d_n^\ast = 0.
\label{eq:thrm-suffi-source-proof-5}
\end{equation}
Lastly, for each $n \in \bbZ^+$, let choose $\phi_n : \cX_n \rightarrow \cY_n$, such that $d(Y_n, \tY_n) \leq 2 d_n^\ast$ and form the sequence of mappings $\{ \phi_n : \cX_n \rightarrow \cY_n\}_{n=1}^\infty$. For this sequence of mappings, we have
\begin{equation}
\limsup_{n \rightarrow \infty} d(Y_n, \tY_n) \leq 2 \limsup_{n \rightarrow \infty} d_n^\ast = 0,
\label{eq:thrm-suffi-source-proof-6}
\end{equation}
where equality follows from \eqref{eq:thrm-suffi-source-proof-5}. Using the definition of variational distance and \eqref{eq:thrm-suffi-source-proof-6}, we have
\begin{equation*}
0 \leq \liminf_{n \rightarrow \infty} d(Y_n, \tY_n) \leq \limsup_{n \rightarrow \infty} d(Y_n, \tY_n) \leq 0,
\end{equation*}
which implies that $\lim_{n \rightarrow \infty} d(Y_n, \tY_n) =0$, which was to be shown.
\end{proof}

\subsection{Source Simulation-Necessary Condition}
\label{ssec:source-approximation-necc-cond}
\begin{theorem}
If $\bX$ is an approximating source for $\bY$, then 
\begin{equation}
\inf_{0 < \epsilon < 1} \liminf_{n \rightarrow \infty} \inf_{0 \leq \delta < 1 - \epsilon} \{ c_n^x(\delta + \epsilon) - c_n^y(\delta) \} \geq 0,
\label{eq:necc1}
\end{equation}
where $c_n^x(\delta)$ and $c_n^y(\delta)$ are as defined in \eqref{eq:defn-c}.
\label{thrm:thrm-source-necc}
\end{theorem}
\begin{proof}

Let $\{ \phi_n \, : \, \cX_n \rightarrow \cY_n\}_{n=1}^\infty$ be a sequence of mappings such that $\lim_{n \rightarrow \infty} d(Y_n, \phi_n(X_n)) = 0$ and $\tY_n \eqdef \phi_n(X_n)$.

We will prove the theorem in two main steps. In the first step, we prove another condition, which essentially states that if $\bX$ is an approximating source for $\bY$, then asymptotically, the cumulative distribution of its entropy spectrum is greater than that of $\bY$. In the second step, we prove that the aforementioned condition implies \eqref{eq:necc1}.

Before stating the proof, we need the following definition:
\begin{definition}
Let $U,V$ be real valued random variables. The L\'{e}vy distance between them, denoted as $L(U,V)$, is defined as
\begin{equation}
L(U,V) \eqdef \inf\left\{ \mu \in \bbR^+ : \forall x \in \bbR, \Pr\left\{ U \leq x - \mu \right\} - \mu \leq \Pr\left\{ V \leq x \right\} \leq \Pr\left\{ U \leq x + \mu \right\} + \mu \right\}.
\label{eq:levy-dist}
\end{equation}
\label{def:levy-dist}
\end{definition}
We continue with the following lemma
\begin{lemma}
For $Y_n, \tY_n \in \cY_n$,
\begin{equation}
\mbox{If } \lim_{n \rightarrow \infty} d\left( Y_n, \tY_n \right) = 0, \mbox{ then } \lim_{n \rightarrow \infty} L \left( \log \frac{1}{P_{Y_n}(Y_n)},  \log \frac{1}{P_{\tY_n}(\tY_n)} \right) = 0.
\label{eq:necc-lemma1}
\end{equation}
\label{lem:necc-lemma1}
\end{lemma}
\begin{proof}

Proof readily follows from the same arguments used in Theorem~2.1.3 of \cite{han03}, of which RHS is \\
 $\lim_{n \rightarrow \infty} L \left( \frac{1}{n}\log \frac{1}{P_{Y_n}(Y_n)}, \frac{1}{n} \log \frac{1}{P_{\tY_n}(\tY_n)} \right) = 0$, and the particular proof does not depend on the existence of $1/n$ factors. 
\end{proof}

Next, we prove the following lemma, which is the entropy spectrum counterpart for the well-known fact \cite{cover91} `any deterministic mapping of a random variable cannot increase its entropy.'
\begin{lemma} 
For any $n \in \bbZ^+$, let $X_n$ be a random variable taking values in $\cX_n$, where $\cX_n$ is countable set and $\tY_n = \phi_n\left(X_n\right)$, where $\phi_n: \cX_n \rightarrow \cY_n$ is any deterministic mapping and $\cY_n$ is countable set. Then, we have
\begin{equation}
\forall c \in \bbR, \; P_{\tY_n}\left( \log \frac{1}{P_{\tY_n}(\tY_n)} < c \right) \geq P_{X_n}\left(  \log \frac{1}{P_{X_n}(X_n)} < c \right). 
\label{eq:necc-lemma2}
\end{equation}
\label{lem:necc-lemma2}
\end{lemma}
\begin{proof}
Define the following sets: 
\begin{align*}
T_n^X(c) & \eqdef \left\{ x \in \cX_n : P_{X_n}(x) \leq e^{-c} \right\}, \\
T_n^{\tilde{Y}}(c) \eqdef & \left\{ y \in \cY^n : P_{\tY_n}(y) \leq e^{-c} \right\}.
\end{align*}
If we can show that
\begin{equation}
 \Pr \left\{ T_n^X(c) \right\} - \Pr\left\{ T_n^{\tilde{Y}}(c) \right\}  \geq 0,
\label{eq:necc-lemma2-proof2}
\end{equation}
for arbitrary choices of $c$ and $n$, then this will conclude the proof.
Now, observe that since $\tilde{Y}_n$ is a deterministic function of $X_n$, we have
\begin{equation}
\forall \, y \in \cY_n, P_{\tY_n}\left( y \right) = P_{X_n}\left( \phi_n^{-1}(y) \right).
\label{eq:necc-lemma2-proof3}
\end{equation}
Using \eqref{eq:necc-lemma2-proof3} and recalling the definition of $T_n^{\tilde{Y}}$, we have
\begin{equation}
\Pr\{ T_n^{\tilde{Y}}(c)\} = P_{X_n}\left( x \in \cX_n : x \in \phi_n^{-1}\left( T_n^{\tilde{Y}}(c) \right)\right).
\label{eq:necc-lemma2-proof3.5}
\end{equation}
Now, if $T_n^{\tilde{Y}}(c) = \emptyset$, then \eqref{eq:necc-lemma2-proof2} holds, hence we are done. Suppose this is not the case. 
Then, we choose any $x \in \phi_n^{-1}\left( T_n^{\tilde{Y}}(c) \right)$. We know that $\phi_n(x) \in T_n^{\tilde{Y}}(c)$, hence $P_{X_n}(x) \leq e^{-c}$ (otherwise $\phi_n(x)$ cannot be an element of $T_n^{\tilde{Y}}$ by recalling the definition of this set and \eqref{eq:necc-lemma2-proof3}), which implies $x \in T_n^X(c)$. Since $x \in  \phi_n^{-1}\left( T_n^{\tilde{Y}}(c) \right)$ is arbitrary, we conclude that 
\begin{equation}
\phi_n^{-1}\left( T_n^{\tilde{Y}}(c) \right) \subseteq T_n^X(c).
\label{eq:necc-lemma2-proof4}
\end{equation}
\eqref{eq:necc-lemma2-proof3.5} and \eqref{eq:necc-lemma2-proof4} immediately implies that $\Pr\{ T_n^X(c)\} \geq \Pr\{  \phi_n^{-1}( T_n^{\tilde{Y}}(c))\} = \Pr\{ T_n^{\tilde{Y}}(c)\}$, hence \eqref{eq:necc-lemma2-proof2} holds. 
\end{proof}

\begin{lemma}
If $\bX$ is an approximating source for $\bY$, then 
\begin{equation}
\inf_{\mu \in \bbR^+} \liminf_{n \rightarrow \infty} \inf_{c \in \bbR} \left\{ P_{Y_n}\left( \log \frac{1}{P_{Y_n}(Y_n)} < c + \mu \right) - P_{X_n}\left( \log \frac{1}{P_{X_n}(X_n)} < c \right)  \right\} \geq 0
\label{eq:necc-lemma3}
\end{equation}
\label{lem:necc-lemma3}
\end{lemma}

\begin{proof}

Suppose $\bX$ approximates $\bY$. Then, using \eqref{eq:necc-lemma1}, we have 
\begin{equation}
\inf_{\mu^\prime \in \bbR^+} \liminf_{n \rightarrow \infty} \inf_{c \in \bbR} P_{Y_n}\left(\log \frac{1}{P_{Y_n}(Y_n)} \leq c + \mu^\prime \right) + \mu^\prime - P_{\tY_n}\left( \log \frac{1}{P_{\tY_n}(\tY_n)} \leq c \right) \geq 0. 
\label{eq:necc-lemma3-proof1}
\end{equation}
Now, observe that we have
\begin{equation}
\forall \mu^\prime \in \bbR^+, \, P_{Y_n}\left( \log \frac{1}{P_{Y_n}(Y_n)} \leq c + \mu^\prime \right) + \mu^\prime \leq P_{Y_n}\left( \log \frac{1}{P_{Y_n}(Y_n)} < c + 2 \mu^\prime \right) + 2 \mu^\prime.
\label{eq:necc-lemma3-proof2}
\end{equation}
Further, 
\begin{equation}
\forall c \in \bbR, \;  P_{\tY_n}\left( \log \frac{1}{P_{\tY_n}(\tY_n)} \leq c \right) \geq  P_{\tY_n}\left( \log \frac{1}{P_{\tY_n}(Y_n)} < c \right) \geq P_{X_n}\left( \log \frac{1}{P_{X_n}(X_n)} < c\right),
\label{eq:necc-lemma3-proof3}
\end{equation}
where second inequality follows recalling \eqref{eq:necc-lemma2}.

Using \eqref{eq:necc-lemma3-proof2} and \eqref{eq:necc-lemma3-proof3} in \eqref{eq:necc-lemma3-proof1} (by defining $\mu \eqdef 2 \mu^\prime$) yields

\begin{equation*}
\inf_{\mu \in \bbR^+} \liminf_{n \rightarrow \infty} \inf_{c \in \bbR} \left\{ P_{Y_n}\left( \log \frac{1}{P_{Y_n}(Y_n)} < c + \mu \right) + \mu - P_{X_n}\left( \log \frac{1}{P_{X_n}(X_n)} < c\right) \right\}\geq 0,
\end{equation*}
which further implies
\begin{eqnarray}
& & \liminf_{\mu \rightarrow 0^+} \liminf_{n \rightarrow \infty} \inf_{c \in \bbR} \left\{P_{Y_n}\left( \log \frac{1}{P_{Y_n}(Y_n)} < c + \mu \right)  - P_{X_n}\left( \log \frac{1}{P_{X_n}(X_n)} < c\right) \right\} = \nonumber \\
& & \liminf_{\mu \rightarrow 0^+} \liminf_{n \rightarrow \infty} \inf_{c \in \bbR} \left\{ P_{Y_n}\left( \log \frac{1}{P_{Y_n}(Y_n)} < c + \mu \right) + \mu - P_{X_n}\left( \log \frac{1}{P_{X_n}(X_n)} < c\right) \right\} \geq 0. \label{eq:necc-lemma3-proof4}
\end{eqnarray}

Now, observe that $P_{Y_n}\left(\log \frac{1}{P_{Y_n}(Y_n)} < c + \mu \right)$ is non-increasing with decreasing $\mu$, therefore \eqref{eq:necc-lemma3-proof4} implies that 
\begin{eqnarray*}
& & \inf_{\mu \in \bbR^+} \liminf_{n \rightarrow \infty} \inf_{c \in \bbR} \left\{ P_{Y_n}\left( \log \frac{1}{P_{Y_n}(Y_n)} < c + \mu \right)  - P_{X_n}\left( \log \frac{1}{P_{X_n}(X_n)} < c \right) \right\}  = \\
& & \liminf_{\mu \rightarrow 0^+} \liminf_{n \rightarrow \infty} \inf_{c \in \bbR} \left\{ P_{Y_n}\left( \log \frac{1}{P_{Y_n}(Y_n)} < c + \mu \right)  - P_{X_n}\left( \log \frac{1}{P_{X_n}(X_n)} < c \right) \right\} \geq  0, 
\end{eqnarray*}
which was to be shown.
\end{proof}

Now we finished the first step of the proof. What remains is to prove the following lemma, which is the second step of the proof.
\begin{lemma}
If
\begin{equation}
\inf_{\mu \in \bbR^+} \liminf_{n \rightarrow \infty} \inf_{c \in \bbR} \left\{ P_{Y_n}\left( \log \frac{1}{P_{Y_n}(Y_n)} < c + \mu \right) - P_{X_n}\left( \log \frac{1}{P_{X_n}(X_n)} < c \right) \right\} \geq 0,
\label{eq:necc-lemma4}
\end{equation}
then we have
\[
\inf_{\epsilon \in (0,1)} \liminf_{n \rightarrow \infty} \inf_{\delta \in [0, 1- \epsilon)} c_n^x(\delta + \epsilon) - c_n^y(\delta) \geq 0. 
\]
\label{lem:necc-lemma4}
\end{lemma}
\begin{proof}

First observe that using \eqref{eq:necc-lemma4}, we have
\begin{equation}
\forall \mu \in \bbR^+, \; \liminf_{n \rightarrow \infty} \inf_{c \in \bbR} \left\{ P_{Y_n}\left( \log \frac{1}{P_{Y_n}(Y_n)} < c + \mu \right) - P_{X_n}\left( \log \frac{1}{P_{X_n}(X_n)} < c \right)  \right\} \geq 0.  
\label{eq:necc-lemma4-proof1}
\end{equation}
Now consider any $\epsilon \in (0,1)$. Next, fix an arbitrary $\mu \in \bbR^+ $ and a sufficiently large $n$, such that we have
\begin{equation}
\inf_{\delta \in (0,1]} \left\{ P_{X_n}\left( \log \frac{1}{P_{X_n}(X_n)} \geq c_n^y(1 - \delta) - \mu \right) -  P_{Y_n}\left( \log \frac{1}{P_{Y_n}(Y_n)} \geq c_n^y( 1- \delta) \right) \right\} \geq - \epsilon.
\label{eq:necc-lemma4-proof2}  
\end{equation}
Consider any $\delta \in (0, 1]$ and observe that we have (cf. definition of $c_n^y(\delta)$) 
\begin{equation}
P_{Y_n}\left( \log \frac{1}{P_{Y_n}(Y_n)} \geq c_n^y( 1- \delta) \right) \geq \delta.
\label{eq:necc-lemma4-proof3}
\end{equation}
Moreover, using \eqref{eq:necc-lemma4-proof2} and \eqref{eq:necc-lemma4-proof3}, we have
\begin{equation}
P_{X_n}\left(\log \frac{1}{P_{X_n}(X_n)} \geq c_n^y( 1- \delta) - \mu \right) \geq \delta - \epsilon.
\label{eq:necc-lemma4-proof4}
\end{equation}
Recalling the definition of $c_n^x(\delta)$, \eqref{eq:necc-lemma4-proof4} immediately implies 
\begin{equation}
c_n^x(1 - \delta + \epsilon) \geq c_n^y(1- \delta) - \mu,
\label{eq:necc-lemma4-proof5}
\end{equation}
for any $\epsilon \in (0,1)$. Since $\mu \in \bbR^+$ is arbitrary, \eqref{eq:necc-lemma4-proof5} implies that we have 
\begin{equation}
\liminf_{n \rightarrow \infty} \inf_{\delta \in (\epsilon, 1]}\{ c_n^x(1 - \delta + \epsilon) - c_n^y(1 - \delta) \}\geq 0.
\label{eq:necc-lemma4-proof6}
\end{equation}
\eqref{eq:necc-lemma4-proof6} yields
\begin{equation}
\liminf_{n \rightarrow \infty} \inf_{\delta \in [0,1- \epsilon)} \{ c_n^x(\delta + \epsilon) - c_n^y(\delta) \} \geq 0.
\label{eq:necc-lemma4-proof7}
\end{equation}

Since $\epsilon \in (0,1)$ is arbitrary, \eqref{eq:necc-lemma4-proof7} yields
\begin{equation*}
\inf_{\epsilon \in (0,1)} \liminf_{n \rightarrow \infty} \inf_{\delta \in [0, 1- \epsilon)} \{ c_n^x(\delta + \epsilon) - c_n^y(\delta) \}\geq 0.
\end{equation*}
which was to be shown.
\end{proof}

Combining Lemma~\ref{lem:necc-lemma3} and \ref{lem:necc-lemma4} we conclude that \eqref{eq:necc1} follows.
\end{proof}

\subsection{Source Simulation-Comparison of the Necessary Condition to Its State-of-the-Art Counterpart}
\label{ssec:source-approximation-comparison}
In this section, we demonstrate that the necessary condition of Theorem~\ref{thrm:thrm-source-necc} is strictly stronger than the necessary condition of \cite{nagaoka96b} (cf. \eqref{eq:japanese-necessary}), which is valid for only \emph{finite alphabets}. First, we prove the following theorem.

\begin{theorem}
If \eqref{eq:necc1} holds, then
\begin{equation}
\forall n, \exists \, P_{X_n, Y_n}, \textrm{ with marginals } P_{X_n}, P_{Y_n}, \textrm{ s.t. }
p-\liminf_{n \rightarrow \infty} \left \{ \frac{1}{n}\log \frac{1}{P_{X_n}(X_n)} - \frac{1}{n}\log \frac{1}{P_{Y_n}(Y_n)} \right\} \geq 0,
\label{eq:thrm-source-necc-comp}
\end{equation}
where the probability measure is $P_{X_n,Y_n}$.
\label{thrm:thrm-source-necc-comp}
\end{theorem}

\begin{proof}

First, we prove the following lemma.
\begin{lemma}
If \eqref{eq:necc1} holds, then
\begin{equation}
\forall n \in \bbZ^+, \exists \{P_{X_n, Y_n}\}_{n \geq 1}, \textrm{ with marginals } P_{X_n}, P_{Y_n}, \textrm{ s.t. } p-\liminf_{n \rightarrow \infty} \left\{ \log \frac{1}{P_{X_n}(X_n)} - \log\frac{1}{P_{Y_n}(Y_n)}\right\} \geq 0.
\label{eq:lem4}  
\end{equation} 
\label{lem:lem4}
\end{lemma}
\begin{proof}

\eqref{eq:necc1} is equivalent to
\begin{equation}
\forall \epsilon \in (0,1), \, \forall \gamma \in \bbR^+, \exists N(\epsilon, \gamma) \in \bbZ^+, \mbox{ s.t. } \forall n \geq N(\epsilon, \gamma), \forall \delta \in [0, 1-\epsilon), \, c_n^x(\delta + \epsilon) - c_n^y(\delta) \geq - \gamma.
\label{eq:lem4-proof1}
\end{equation}
\eqref{eq:lem4-proof1} implies that
\begin{equation}
\forall \epsilon \in (0,1), \, \forall \gamma \in \bbR^+, \exists N(\epsilon, \gamma) \in \bbZ^+, \mbox{ s.t. } \forall n \geq N(\epsilon, \gamma), \, \mu(\delta \in [0, 1-\epsilon) : c_n^x(\delta + \epsilon) -c_n^y(\delta) < -\gamma) = 0.
\label{eq:lem4-proof2}
\end{equation}
Consider any $\epsilon \in (0,1) \mbox{ and } \gamma \in \bbR^+$ and fix some $n \geq N(\epsilon, \gamma)$. Next, define $\Omega \sim U[0,1]$ and
\begin{equation*}
\tilde{\Omega}(\omega) = \left\{ \begin{array}{cl} \omega + \epsilon, & \mbox{ if } \omega < 1- \epsilon,  \\
\omega + \epsilon -1 , & \mbox{ if } \omega \geq  1- \epsilon 
\end{array} \right.
\end{equation*}
Let $\cS_n(X), \, \Delta^x$ and $c_n^x(\delta)$ (resp. $\cS_n(Y), \, \Delta^y$ and $c_n^y(\delta)$) denote the quantities defined in \eqref{eq:defn-sz}, \eqref{eq:defn-c-partition} and \eqref{eq:defn-c}, respectively, for $\cX_n$ (resp. $\cY_n$). Using the $\Omega, \tilde{\Omega}$ pair, define the following joint distribution for $\bX^n, \bY^n$
\begin{equation}
P_{X_n, Y_n}(x_i, y_i) \eqdef  \Pr \left\{ c_n^x(\tilde{\Omega}) = \log \frac{1}{\delta_i^x - \delta_{i-1}^x}, c_n^y(\Omega) = \log \frac{1}{\delta_j^y - \delta_{j-1}^y}\right\},
\label{eq:lem4-proof3}
\end{equation}
for any $(i,j) \in \{ 1, \ldots, |\cS_n(X)| \} \times \{ 1, \ldots, |\cS_n(Y)| \}$.
Note that as a direct consequence of \eqref{eq:lem4-proof3}, the marginals of $P_{X_n, Y_n}$ are $P_{X_n}$ and $P_{Y_n}$. Further, we have
\begin{align}
P_{X_n, Y_n}\left\{ \log \frac{1}{P_{X_n}(X_n)} - \log \frac{1}{P_{Y_n}(Y_n)} < - \gamma \right\} & =  \Pr\left\{ c_n^x(\tilde{\Omega}) - c_n^y(\Omega) < - \gamma\right\} , \label{eq:lem4-proof4} \\
 & \leq  \Pr \left\{ \omega \in [0, 1- \epsilon) : c_n^x(\omega + \epsilon) - c_n^y(\omega) < - \gamma \right\} + \epsilon, \nonumber \\
 & \leq  \mu(\delta \in  [0, 1- \epsilon) : c_n^x(\delta + \epsilon) - c_n^y(\delta) < -\gamma) + \epsilon, \label{eq:lem4-proof5} \\
 & \leq \epsilon, \label{eq:lem4-proof6}
\end{align}
where \eqref{eq:lem4-proof4} follows from \eqref{eq:lem4-proof3}, \eqref{eq:lem4-proof5} follows from the definition of $\Omega$ and $\tilde{\Omega}$, and \eqref{eq:lem4-proof6} follows from \eqref{eq:lem4-proof2}.

Since $\epsilon \in (0,1), \, \gamma \in \bbR^+$ and $n \geq N(\epsilon, \gamma)$ is arbitrary, \eqref{eq:lem4-proof6} implies that
\begin{equation}
\forall \gamma \in \bbR^+, \lim_{n \rightarrow \infty} P_{X_n, Y_n}\left(\log \frac{1}{P_{X_n}(X_n)} - \log \frac{1}{P_{Y_n}(Y_n)} < - \gamma \right) = 0.
\label{eq:lem4-proof7}
\end{equation}
Recalling definition of limit infimum in probability, \eqref{eq:lem4-proof7} implies that for this particular construction of $P_{X_n, Y_n}$, we have
\[
p-\liminf_{n \rightarrow \infty} \left\{ \log\frac{1}{P_{X_n}(X_n)} - \log\frac{1}{P_{Y_n}(Y_n)}\right\} \geq 0,
\]
where probability measure is $P_{X_n,Y_n}$, which was to be shown.
\end{proof}

Next, we conclude the proof. Lemma~\ref{lem:lem4} directly implies that if \eqref{eq:necc1} holds, then $\exists \{P_{X_n,Y_n}\}_{n \geq 1}$ with marginals $P_{X_n} \mbox{ and } P_{Y_n}$, such that
\begin{equation}
\forall \epsilon, \gamma \in \bbR^+, \exists N(\epsilon, \gamma) \in \bbZ^+, \mbox{ s.t. } \forall n \geq N(\epsilon, \gamma), P_{X_n, Y_n}\left(\frac{1}{n}\log\frac{1}{P_{X_n}(X_n)} - \frac{1}{n}\log\frac{1}{P_{Y_n}(Y_n)} < -\frac{\gamma}{n}\right) \leq \epsilon. 
\label{eq:thrm-3-proof1}
\end{equation} 
\eqref{eq:thrm-3-proof1} immediately implies that for this sequence of joint distributions, we have
\[
p-\liminf_{n \rightarrow \infty} \left\{ \frac{1}{n} \log\frac{1}{P_{X_n}(X_n)} - \frac{1}{n}\log\frac{1}{P_{Y_n}(Y_n)} \right\} \geq 0,
\]
where probability measure is $P_{X_n,Y_n}$. Hence we are done.
\end{proof}

\begin{remark}
If we take $\cX_n$ (resp. $\cY_n$) as a finite set for all $n \in \bbZ^+$, then \eqref{eq:thrm-source-necc-comp} implies the necessary condition of \cite{nagaoka96b}, hence Theorem~\ref{thrm:thrm-source-necc-comp} implies that our necessary condition is at least as good as the one in \cite{nagaoka96b}. Moreover, recalling Example~4 stated in Section~\ref{ssec:examples} the converse is \emph{not} true, in other words \eqref{eq:japanese-necessary} does \emph{not} imply our necessary condition. Therefore, we conclude that our necessary condition for the source simulation problem is strictly stronger than its state-of-the-art counterpart. 
\label{rem:remark-necc-cond-comp2}
\end{remark}


\section{Channel Simulation}
\label{sec:channel-approximation}
Throughout this section, $\bX = \{ X_n\}_{n=1}^\infty$ and $\bZ = \{ Z_n\}_{n=1}^\infty$ denote two general sources, where for all $n$, $X_n$ is a random variable taking values in $\cX_n$, such that $\cX_n$ is countable set  (resp. $Z_n$ is a random variable taking values in $\cZ_n$, such that $\cZ_n$ is countable set) with an arbitrary coupling $P_{Z_n|X_n}$. Further, $\bW_{Y|X} = \left\{ W_{Y_n|X_n}(Y_n|X_n) \right\}_{n=1}^{\infty}$ denotes a general channel, where for all $n \in \bbZ^+$, $W_{Y_n|X_n}$ denotes a conditional pmf over $\cY_n \times \cX_n$, with $\cY_n$ being countable set.


\begin{theorem}
If 
\begin{equation} 
\forall \gamma \in \bbR, \,  \lim_{n \rightarrow \infty} \mbox{E}_{P_{X_n}}\left[\mu( \delta \in [0,1) \, : \, c_n^{z|x}(\delta, X_n) - c_n^w(\delta, X_n) < \gamma) \right] = 0,
\label{eq:thrm-channel-suff}
\end{equation}
then $\bZ$ is an approximating source for $\bW$, given $\bX$. \\
\label{thrm:thrm-channel-suff}
\end{theorem}

\begin{proof}

First, observe that \eqref{eq:thrm-channel-suff} is equivalent to
\begin{equation}
\forall \epsilon \in \bbR^+, \gamma \in \bbR, \, \exists N(\epsilon, \gamma) \in \bbZ^+, \mbox{ s.t. } \forall n \geq N(\epsilon, \gamma), \mbox{E}_{P_{X_n}}\left[\mu( \delta \in [0,1) \, : \, c_n^{z|x}(\delta, X_n) - c_n^w(\delta, X_n) < \gamma) \right] \leq \epsilon. 
\label{eq:thrm-channel-suff-proof1}
\end{equation}
Next, consider any $\epsilon \in (0,1)$, $\gamma \in \bbR^+$, such that $e^{- \gamma} \leq \epsilon$ and fix some $n \geq N(\epsilon, \gamma)$. Now, observe that for any $x \in \cX_n$, both $Z_n$ and $Y_n$ are general sources with distributions $P_{Z_n|X_n}(\cdot|x)$ and $W_{Y_n|X_n}(\cdot|x)$. For any $x \in \cX_n$ (by defining $\cE^n(\gamma, x) \eqdef \{ \delta \in [0,1) : c_n^{x|z}(\delta, x) - c_n^w(\delta, x) <\gamma \}$) Proposition~\ref{prop:single-source-prop-1} guarantees that we have
\begin{equation}
\exists \phi_n^{x} : \cZ_n \rightarrow \cY_n, \mbox{ with } d_{x} \leq 9 \epsilon + 10 \mu(\cE^n(\gamma, x)),
\label{eq:thrm-channel-suff-proof2}
\end{equation}
where 
\begin{equation}
\forall x \in \cX_n, \, d_{x} \eqdef \sum_{y \in \cY_n} \left|W_{Y_n|X_n}(y|x) - P_{Z_n|X_n}( (\phi_n^{x})^{-1}(y) | x)\right|.
\label{eq:thrm-channel-suff-proof3}
\end{equation}
Using $\phi_n^{x}$, we define the following mapping
\begin{equation}
\varphi_n : \cX_n \times \cZ_n \rightarrow \cY_n, \mbox{ s.t. } \forall (x, z) \in \cX_n \times \cZ_n, \varphi(x, z) = \phi_n^{x}(z).
\label{eq:thrm-channel-suff-proof4}
\end{equation}
We have 
\begin{align}
d(X_nY_n, X_n \varphi_n(X_n,Z_n))  & =  \mbox{E}_{P_{X_n}}\left[ d_{X_n}\right], \label{eq:thrm-channel-suff-proof5} \\
& \leq  19 \epsilon, \label{eq:thrm-channel-suff-proof6}
\end{align}
where \eqref{eq:thrm-channel-suff-proof5} follows from \eqref{eq:thrm-channel-suff-proof3}, \eqref{eq:thrm-channel-suff-proof4} and recalling definition of variational distance and \eqref{eq:thrm-channel-suff-proof6} follows from \eqref{eq:thrm-channel-suff-proof1} and \eqref{eq:thrm-channel-suff-proof2}.

Using arguments similar to those of the proof of Theorem~\ref{thrm:thrm-suffi-source}, one can conclude the proof.

\end{proof}


\begin{theorem}
If $\bZ$ is an approximating source for $\bW$, given $\bX$, then 
\begin{equation}
\forall \epsilon \in (0,1), \, \forall \gamma \in \bbR^+, \, \lim_{n \rightarrow \infty} \mbox{E}_{P_{X_n}}\left[\mu( \delta \in [0,1-\epsilon) \, : \, c_n^{z|x}(\delta + \epsilon, X_n) - c_n^w(\delta, X_n) < -\gamma) \right] = 0.
\label{eq:thrm-channel-necc}
\end{equation}
\label{thrm:thrm-channel-necc}
\end{theorem}

\begin{proof}
Let $\{ \varphi_n \, : \, \cX_n \times \cZ_n \rightarrow \cY_n\}_{n=1}^\infty$ be a sequence of mappings with 
\begin{equation}
\lim_{n \rightarrow \infty} d(X_nY_n, X_n\varphi_n(X_n, Z_n)) =0.
\label{eq:thrm-channel-necc-proof0.5} 
\end{equation}
Let $N \in \bbZ^+$ be such that $\forall n \geq N$, $\mbox{E}_{P_{X_n}}[d_{X_n}] < 1$, where $d_{X_n}$ is as defined in \eqref{eq:thrm-channel-suff-proof3}. For the sake of notational convenience, we define
\begin{equation}
\tilde{\mu}(\epsilon, \gamma, x_n) \eqdef \mu\left( \delta \in [0,1-\epsilon) \, : \, c_n^{z|x}(\delta + \epsilon, x_n) - c_n^w(\delta) < - \gamma \right),
\label{eq:thrm-channel-necc-proof1}
\end{equation}
for any $\epsilon \in (0,1)$, $\gamma \in \bbR^+$ and $x_n \in \cX_n$. Next, we consider any $n \geq N$ and define the following set:
\begin{equation}
\cS_n \eqdef \{ x_n \in \cX_n \, : \, d_{X_n} \geq (\mbox{E}_{P_{X_n}}[d_{X_n}])^{1/2}\}.
\label{eq:thrm-channel-necc-proof2}
\end{equation}
Using Markov's inequality, \eqref{eq:thrm-channel-necc-proof2} implies that
\begin{equation}
\Pr \left\{ X_n \in \cS_n \right\} \leq (\mbox{E}_{P_{X_n}}[d_{X_n}])^{1/2}.
\label{eq:thrm-channel-necc-proof3}
\end{equation}
Moreover, define $\tilde{x} = \{\tilde{x}_n\}_{n=1}^\infty$, where $\forall n \in \bbZ^+, \, \tilde{x}_n \in \cX_n$  and for all $n \geq N$, $\tilde{x}_n \in \cS_n^c$ and $\tilde{\mu}(\epsilon, \gamma, \tilde{x}_n) \geq \frac{1}{2} \sup_{x_n \in \cS_n^c} \tilde{\mu}(\epsilon, \gamma, x_n)$. Observe that \eqref{eq:thrm-channel-necc-proof3} guarantees the existence of such a sequence, since $\cS_n^c \neq \varnothing$ for all $n \geq N$. By the definition of $\tilde{x}$, we have $\lim_{n \rightarrow \infty} d_{\tilde{x}_n} =0$, hence the necessary condition of source simulation, i.e. \eqref{eq:necc1}, implies that
\begin{equation}
\lim_{n \rightarrow \infty} \tilde{\mu}(\epsilon, \gamma, \tilde{x}_n) = 0,
\label{eq:thrm-channel-necc-proof3.5}
\end{equation}
for any $\epsilon \in (0,1)$ and $\gamma \in \bbR^+$.

Consider any $n \geq N$, $\epsilon \in (0,1)$ and $\gamma \in \bbR^+$. Using the law of total expectation, we have
\begin{align}
\mbox{E}_{P_{X_n}}[\tilde{\mu}(\epsilon, \gamma, X_n)]  & = \mbox{E}[\tilde{\mu}(\epsilon, \gamma, X_n) \, | \, X_n \in \cS_n] \cdot \Pr\left\{ X_n \in \cS_n \right\} +  \mbox{E}[\tilde{\mu}(\epsilon, \gamma, X_n) \, | \, X_n \in \cS_n^c] \cdot \Pr \left\{ X_n \in \cS_n^c \right\}, \nonumber \\
 & \leq (\mbox{E}_{P_{X_n}}[d_{X_n}])^{1/2} +  \mbox{E}[\tilde{\mu}(\epsilon, \gamma, X_n) \, | \, X_n \in \cS_n^c], \label{eq:thrm-channel-necc-proof4}\\
 & \leq (\mbox{E}_{P_{X_n}}[d_{X_n}])^{1/2} + 2 \tilde{\mu}(\epsilon, \gamma, \tilde{x}_n), \label{eq:thrm-channel-necc-proof5}
\end{align} 
where \eqref{eq:thrm-channel-necc-proof4} follows from \eqref{eq:thrm-channel-necc-proof3} and \eqref{eq:thrm-channel-necc-proof5} follows from the definition of $\tilde{x}$. \eqref{eq:thrm-channel-necc-proof5} implies that
\begin{align}
\limsup_{n \rightarrow \infty}\mbox{E}_{P_{X_n}}[\tilde{\mu}(\epsilon, \gamma, X_n)]  & \leq \lim_{n \rightarrow \infty} (\mbox{E}_{P_{X_n}}[d_{X_n}])^{1/2} + 2 \lim_{n \rightarrow \infty} \tilde{\mu}(\epsilon, \gamma, \tilde{x}_n), \nonumber \\
 & \leq 0, \label{eq:thrm-channel-necc-proof6}
\end{align}
where \eqref{eq:thrm-channel-necc-proof6} follows from \eqref{eq:thrm-channel-necc-proof0.5} and \eqref{eq:thrm-channel-necc-proof3.5}. Since $\mbox{E}_{P_{X_n}}[\tilde{\mu}(\epsilon, \gamma, X_n)]  \geq 0$, $\forall n \in \bbZ^+$ and $\epsilon \in (0,1)$, $\gamma \in \bbR^+$ are arbitrary, \eqref{eq:thrm-channel-necc-proof6} implies \eqref{eq:thrm-channel-necc}.
\end{proof}

\section{Conclusion}
\label{sec:conclusion}
In this paper, we consider source and channel simulation problems for the general case and prove \emph{essentially the same} necessary and sufficient conditions. The necessary and sufficient conditions for the source (resp. channel) simulation problems exploits the knowledge of the \emph{whole} entropy (resp. conditional entropy) density of the target source (resp. general channel) and the coin source. Moreover, our necessary condition for the source simulation problem is strictly stronger than its state-of-the-art counter part (cf. \cite{nagaoka96b})
which is valid for only finite alphabets. As a future research problem, this kind of approach may also be exploited to solve the general case of the \emph{approximation theory of output statistics} (which is originally formulated in \cite{verdu93} for the special case of fair coin flips as the coin source) problem which is still an open problem (cf. \cite{karthik00}).

\end{document}